\documentclass[sigconf, nonacm, review=true]{acmart}

\usepackage[utf8]{inputenc} 
\usepackage{amsmath}
	\usepackage{amsfonts}
\usepackage{mathtools}
\usepackage{graphicx} %
\usepackage{url}
\usepackage{enumitem} %

\usepackage{dsfont} %

 \usepackage{bm} %
\usepackage{stmaryrd} %

\usepackage{upgreek} %
\usepackage{bbding} %

\usepackage{listings} %

\usepackage{tabularx} %
\usepackage{booktabs} %
\usepackage{multirow}
\usepackage{hhline} %
\usepackage{diagbox}

\usepackage{xcolor,colortbl}    %

\usepackage{tikz} %
\usetikzlibrary{matrix}
\usetikzlibrary{calc}
\usetikzlibrary{math}
\usetikzlibrary{arrows.meta,positioning}

\usetikzlibrary{positioning}
\newcommand*\circled[1]{\tikz[baseline=(char.base)]{
            \node[shape=circle,fill=white,draw,text=black,inner sep=1.2pt] (char) {#1};}}

\usepackage{easybmat} %

\usepackage{adjustbox}
\def\eox{\unskip\kern 10pt{\unitlength1pt\linethickness{.4pt}$\diamondsuit${}}} %

\usepackage{rotating}		%

\usepackage{xspace} %

\usepackage{subcaption} %
\newcommand{\wolf}[1]{{{\color{magenta}{[\textbf{WG}: #1]}}}}
\newcommand{\agapi}[1]{{{\color{cyan}{[\textbf{AG}: #1]}}}}
\newcommand{\benny}[1]{{{\color{purple}{[\textbf{BK}: #1]}}}}
\newcommand{\ilias}[1]{{{\color{orange}{[\textbf{IF}: #1]}}}}
\newcommand{\note}[1]{{{\color{green}{[#1]}}}} %

\newcommand{\rev}[2]{{{\color{red}{[**Reviewer#1**: #2]}}}} %

\newcommand{\anonymize}[0]{\textcolor{red}{[ANONYMIZED] }}

\newcommand{\hide}[1]{} 
\newcommand{\hidetwo}[2]{} 
\newcommand{\mkclean}{
    \renewcommand{\note}{\hide}
    \renewcommand{\wolf}{\hide}
    \renewcommand{\agapi}{\hide}
    \renewcommand{\benny}{\hide}
    \renewcommand{\ilias}{\hide}

    \renewcommand{\rev}{\hidetwo}

    \renewcommand{\anonymize}{\hide}
}

\mkclean %

\usepackage[ruled,noend,linesnumbered]{algorithm2e} %
\DontPrintSemicolon

\usepackage[capitalise,nameinlink]{cleveref} %

\crefname{algocf}{alg.}{algs.}
\Crefname{algocf}{Algorithm}{Algorithms}
\crefalias{AlgoLine}{line}%
\crefname{proposition}{prop.}{prop.}
\crefname{figure}{Figure}{Figures}
\crefname{observation}{Observation}{Observations}

	\crefformat{equation}{Eq.~(#2#1#3)}						%
	\crefrangeformat{equation}{(#3#1#4) to~(#5#2#6)}
	\crefmultiformat{equation}{(#2#1#3)}%
	{ and~(#2#1#3)}{, (#2#1#3)}{ and~(#2#1#3)}
	
	\Crefformat{equation}{Equation~(#2#1#3)}						%

\usepackage{aliascnt}

\newaliascnt{corollary}{theorem}

\aliascntresetthe{corollary}

\newaliascnt{example}{theorem}
\newtheorem{example}[example]{Example}
\aliascntresetthe{example}

\newaliascnt{definition}{theorem}
\newtheorem{definition}[definition]{Definition}
\aliascntresetthe{definition}

\newaliascnt{proposition}{theorem}

\aliascntresetthe{proposition}

\newaliascnt{lemma}{theorem}

\aliascntresetthe{lemma}

\newaliascnt{conjecture}{theorem}

\aliascntresetthe{conjecture}

\newtheorem{questionW}{Question}
\newtheorem{resultW}{Result}

{\end{itshape}
}
\usepackage{tcolorbox}		%
\tcbuselibrary{breakable,skins}		%

\tcbset{examplestyle/.style={
		enhanced jigsaw,	%
		colback=blue!08,	%
		colframe=blue!08,	%
		arc=0mm,
		boxrule=0pt,		%
		left=1mm,
		right=1mm,
		left skip= 0mm,  %
		right skip= 0mm, %
		top=1mm,		%
		bottom=1mm,		%
		breakable,		%
		parbox = false,		%
		before={\par\pagebreak[0]\vspace{1mm}\parindent=0pt},		
		after={\par\pagebreak[0]\vspace{1mm}\parindent=0pt},				
		bottomrule = 0mm,
		boxsep = 0mm,					%
		topsep at break=0pt,			%
		bottomsep at break=0pt,			%
		pad at break=0mm,
		pad before break=1mm,		
		pad after break=1mm,		
		bottomrule at break=0mm,
		toprule at break=0mm,		
		}}

\usepackage{footnote}

\tcolorboxenvironment{example}{examplestyle}
\usepackage{footnote}%

\makeatletter
\DeclareRobustCommand*\uell{\mathpalette\@uell\relax}
\newcommand*\@uell[2]{
  \setbox0=\hbox{$#1\ell$}
  \setbox1=\hbox{\rotatebox{10}{$#1\ell$}}
  \dimen0=\wd0 \advance\dimen0 by -\wd1 \divide\dimen0 by 2
  \mathord{\lower 0.1ex \hbox{\kern\dimen0\unhbox1\kern\dimen0}}
}
\makeatother

\usepackage{marginnote}
\newcommand{\rebuttal}[2]{#2}

\AtEndPreamble{%
    \hypersetup{colorlinks,
      linkcolor=purple,
      citecolor=blue,
      urlcolor=ACMDarkBlue,
      filecolor=ACMDarkBlue}}

\definecolor{mygreen}{rgb}{0,0.6,0}
\definecolor{mygray}{rgb}{0.5,0.5,0.5}
\definecolor{mymauve}{rgb}{0.58,0,0.82}
\definecolor{myred}{rgb}{1,0,0}
\definecolor{mylightblue}{rgb}{0.953, 0.957, 0.980} 	%
\definecolor{mylightblueborder}{rgb}{0.567, 0.627, 0.757} 	%
\definecolor{mylightyellow}{rgb}{0.988, 0.957, 0.863} 	%

\RequirePackage{scalerel}
\RequirePackage{tikz}
\usetikzlibrary{svg.path}

\definecolor{orcidlogocol}{HTML}{A6CE39}
\tikzset{
  orcidlogo/.pic={
    \fill[orcidlogocol] svg{M256,128c0,70.7-57.3,128-128,128C57.3,256,0,198.7,0,128C0,57.3,57.3,0,128,0C198.7,0,256,57.3,256,128z};
    \fill[white] svg{M86.3,186.2H70.9V79.1h15.4v48.4V186.2z}
                 svg{M108.9,79.1h41.6c39.6,0,57,28.3,57,53.6c0,27.5-21.5,53.6-56.8,53.6h-41.8V79.1z M124.3,172.4h24.5c34.9,0,42.9-26.5,42.9-39.7c0-21.5-13.7-39.7-43.7-39.7h-23.7V172.4z}
                 svg{M88.7,56.8c0,5.5-4.5,10.1-10.1,10.1c-5.6,0-10.1-4.6-10.1-10.1c0-5.6,4.5-10.1,10.1-10.1C84.2,46.7,88.7,51.3,88.7,56.8z};
  }
}

\RequirePackage{etoolbox}
\DeclareRobustCommand\orcidicon[1]{\href{https://orcid.org/#1}{\mbox{\scalerel*{
\begin{tikzpicture}[yscale=-1, transform shape]
    \pic{orcidlogo};
\end{tikzpicture}
}{|}}}}

\def\e#1{\emph{#1}}
\def\scs{\mathcal{S}}
\def\scr{\mathcal{R}}
\def\dom{\mathsf{dom}}
\def\rels{\mathsf{rel}}

\def\att{\mathsf{att}}
\def\key{\mathsf{key}}

\def\upd{\mathsf{UPD}}
\def\agg{\mathsf{AGG}}
\def\mlp{\mathsf{MLP}}
\def\enc{\mathsf{Enc}}

\def\fks{\mathsf{FK}}
\def\set#1{\mathord{\{#1\}}}
\def\reals{\mathbb{R}}
\def\exlang{\mathbf{EL}}
\newcommand{\defeq}{\vcentcolon=}

\def\cost{\mathsf{cost}}
\def\angs#1{\mathord{\langle #1\rangle}}

\def\devi{\mathsf{dev}}

\newcommand*{\ldblbrace}{\{\mskip-5mu\{}
\newcommand*{\rdblbrace}{\}\mskip-5mu\}}
\def\bag#1{\mathord{\ldblbrace#1\rdblbrace}}

\def\G{\mathord{\mathcal{G}}}
\def\X{\mathord{\mathbf{X}}}

\def\N{\mathcal{N}}
\def\A{\mathcal{A}}
\def\vin{{\mathsf{in}}}
\def\vout{{\mathsf{out}}}
\def\src{{\mathsf{src}}}
\def\tgt{{\mathsf{tgt}}}

\renewcommand{\vec}[1]{\mathord{\mathbf{#1}}}    %

\newcommand{\sql}[1]{\textsf{\textup{#1}}}

\def\tups{\mathsf{Tups}}
\def\attrs{\mathsf{Attr}}
\def\fkeys{\mathsf{FK}}

\newcommand{\introparagraph}[1]{\textbf{#1.}} %

\newcommand{\dist}{\textrm{dist}}

\newcommand{\colmask}{\texttt{Column Mask}}
\newcommand{\fkpkmask}{\texttt{FKPK Mask}}
\newcommand{\filtermask}{\texttt{Filter Mask}}
\newcommand{\pfi}{\texttt{PFI}}
\newcommand{\greedysubset}{\texttt{Local Impact}}
\newcommand{\greedysubsetiterative}{\texttt{Greedy}}
\newcommand{\greedyexpansion}{\texttt{Greedy Expansion}}
\newcommand{\emptymethod}{\texttt{Empty}}
\newcommand{\randomsubset}{\texttt{Random Subset}}

\def\proj{\textsc{Projection}}
\def\join{\textsc{FKJoin}}
\def\select{\textsc{Selection}}
\def\joinproj{\textsc{FKJoinProj}}
\def\projselect{\textsc{ProjSelect}}
\def\joinprojselect{\textsc{FKJoinProjSelect}}

\usepackage{makecell}
\usepackage{multirow}

\makeatletter
\newcommand{\ACMdisablelinenumbers}{%
  \let\ACM@orig@linecountL\ACM@linecountL
  \let\ACM@orig@linecountR\ACM@linecountR
  \let\ACM@linecountL\relax
  \let\ACM@linecountR\relax
}
\newcommand{\ACMenablelinenumbers}{%
  \let\ACM@linecountL\ACM@orig@linecountL
  \let\ACM@linecountR\ACM@orig@linecountR
}

\begin{document}

\title{Database Views as Explanations for Relational Deep Learning}

\author{Agapi Rissaki}
\orcid{0009-0000-6161-6288}
\affiliation{%
    \orcidicon{0009-0000-6161-6288}
    \institution{Northeastern University\country{USA}}
}

\author{Ilias Fountalis}
\orcid{0009-0008-6917-5702}
\affiliation{%
    \orcidicon{0009-0008-6917-5702}
    \institution{RelationalAI\country{USA}}
}

\author{Wolfgang Gatterbauer}
\orcid{0000-0002-9614-0504}
\affiliation{%
    \orcidicon{0000-0002-9614-0504}
    \institution{Northeastern University\country{USA}}
}

\author{Benny Kimelfeld}
\orcid{0000-0002-7156-1572}
\affiliation{%
    \orcidicon{0000-0002-7156-1572}
    \institution{Technion \& RelationalAI\country{Israel}}
}

\begin{abstract}

In recent years, there has been significant progress in the development of deep learning models over relational databases, including architectures based on heterogeneous graph neural networks (hetero-GNNs) and heterogeneous graph transformers. In effect, such architectures state how the database records and links (e.g., foreign-key references) translate into a large, complex numerical expression, involving numerous learnable parameters. This complexity makes it hard to explain, in human-understandable terms, how a model uses the available data to arrive at a given prediction. 

We present a novel framework for explaining machine-learning models over relational databases, where \emph{explanations are view definitions} that highlight focused parts of the database that mostly contribute to the model's prediction. We establish such global abductive explanations by adapting the classic notion of determinacy by Nash, Segoufin, and Vianu (2010). 
In addition to tuning the tradeoff between determinacy and conciseness, the framework allows controlling the level of granularity by adopting different fragments of view definitions, such as ones highlighting whole columns, foreign keys between tables, relevant groups of tuples, and so on. 

\rebuttal{R1D5, R2W2 and R2D}{
We investigate the realization of the framework in the case of hetero-GNNs, and develop a model-specific approach via the notion of learnable masks. For comparison, we propose model-agnostic heuristic baselines and show that our approach is both more efficient and achieves better explanation quality in most cases.} %
Our extensive empirical evaluation on the RelBench collection across diverse domains and record-level tasks demonstrates both the usefulness of our explanations and the efficiency of their generation.

\end{abstract}

\maketitle

\ACMdisablelinenumbers

\section{Introduction}

Machine learning over relations was traditionally done through a reduction to learning over unstructured data: flatten the data into a \e{feature matrix}, with a collection of predefined queries that produce features for every prediction instance, and then learn a flat model~\cite{DBLP:conf/kdd/PerlichP03,DBLP:journals/sigmod/KimelfeldR18,bakshi2021optimized}. Progress in deep learning, and specifically Graph Neural Networks (GNNs)~\cite{wu2020comprehensive} and graph transformers~\cite{DBLP:journals/tmlr/Muller00R24} changed this practice: the best performing models now deploy prediction models directly from the structured data without ad-hoc feature extraction~\cite{chen2025relgnn,4dbinfer_wang2024}. 

Specifically, deep learning is commonly applied to relational databases by translating them into a graph representation:
nodes represent database entities (usually tuples),
and edges represent relationships (usually foreign-primary key constraints).
GNNs or graph transformers are then trained on these graphs~\cite{fey2024position,DBLP:conf/sigmod/CappuzzoPT20,DBLP:conf/icde/TonshoffFGK23,DBLP:conf/eacl/SogaardS17}.
To handle the varying \emph{types} of nodes and edges determined by the database schema, 
\emph{heterogeneous} graph models are used~\cite{DBLP:conf/aaai/ZhaoWSHSY21,DBLP:conf/www/WangJSWYCY19,chen2025relgnn,4dbinfer_wang2024}. 
This creates a two-paradigm pipeline: the relational database defines the graph, while the (graph-based) machine learning library learns the model. 
At inference time, new data is exported to the graph representation, predictions are computed from the learned model, and results are reintegrated into the database.

\begin{figure}[t]
    \centering
    \includegraphics[scale=0.35]{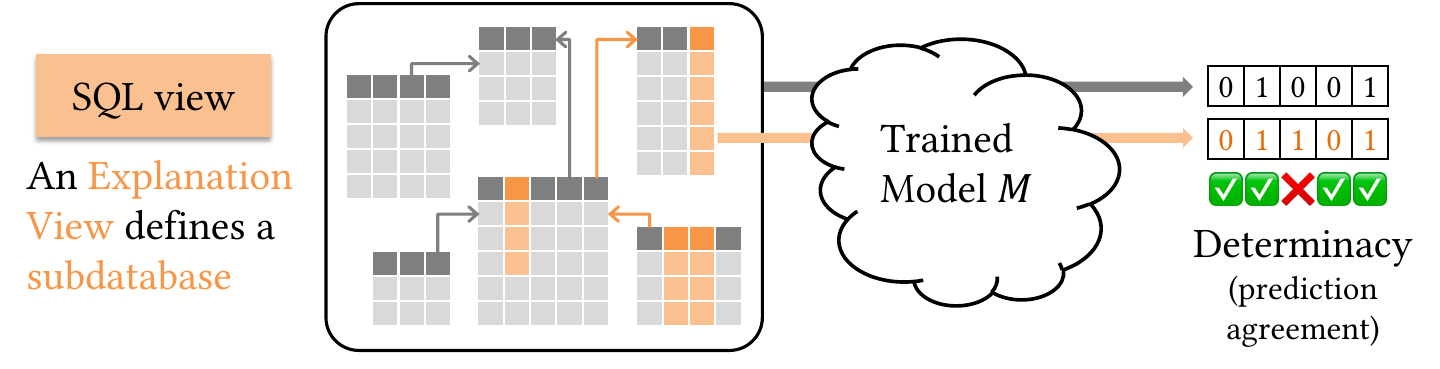}
    \caption{An \emph{explanation view} defines a subdatabase (or more generally, a derived view).
    The rest of the database gets randomly perturbed. As long as the view is kept the same, 
    a trained black-box model gives similar predictions over the perturbed database as the original database.
    We say that the view \emph{soft determines} the prediction.
    }
    \label{Fig_intro}
\end{figure}

The deep model induced by the above graph ML approach over the database can be viewed as a complex numerical function constructed from the database's tuples and values, and parameterized by 
learned weights. 
Consequently, 
the model's behavior is inherently opaque and cannot be readily understood through direct inspection. To address this, the field of \emph{Explainable Artificial Intelligence (XAI)} provides tools that shed light on different aspects of the model~\cite{DBLP:journals/widm/AngelovSJAA21, 10.1145/3635138.3654762/Arenas, marques2023logic, Darwiche-LogicforExplainable}. These include \emph{local explanations}, which account for the model's prediction on a specific instance, and \emph{global explanations}, which aim to characterize the model's general behavior. 
Explanations may be \emph{contrastive} (or counterfactual~\cite{barcelo2025explaining}), identifying hypothetical changes that would alter the model's outcome, or \emph{abductive}, pointing to focused parts of the input that suffice to determine the outcome~\cite{marques2023logic}. 

In this work, we study \emph{global abductive explanations} for deep relational models. Our focus is on settings where the underlying schema consists of multiple interrelated tables with numerous attributes---a common scenario in real-world databases. 
We aim to identify concise subsets of the data that best account for the model's predictions. 
Specifically, we address the question: \emph{Which parts of the database does the model rely on most when making predictions?} 
Hence, our goal is to provide global insights into the model's behavior by uncovering the database fragments that are most influential across multiple predictions.

\begin{example}\label{example:intro}
\rebuttal{R3C1 and R3D1}{
To illustrate the nature of our explanations, consider a relational database with schema \sql{R(rid, $a_1$, $\ldots$, $a_{5}$), S(sid, rid, $b_1$, $\ldots$, $b_5$, flag), T(tid, rid, $c_1$, $\ldots$, $c_5$)}, where \sql{rid} is the primary key of $R$ and a foreign key in both $S$ and $T$. All attributes $a_i, b_i, c_i$ are numerical, while \sql{flag} is boolean. 
Consider two binary classification tasks predicting labels for $R$-records. The ground truth of Task $1$ corresponds to whether at least $3$ linked $S$-records have  \sql{flag = TRUE}, while Task $2$ requires only at least $3$ linked $S$-records (regardless of \sql{flag} values). Task $2$ thus depends purely on the join structure (\sql{on R.rid = S.rid}), while Task $1$ additionally depends on a specific attribute (i.e., the \sql{flag}).

Now suppose a data scientist has trained a relational deep learning model that achieves high accuracy (ROC-AUC > 0.95) on both tasks. Applying our approach, the data scientist can understand what drives each model's predictions. 
For \emph{Task 1}, explanations focused on important features reveal that only \sql{S.flag} is critical, while all other feature columns ($a_i, b_i, c_i$) are excluded. 
To verify whether the relational structure also contributes, the data scientist requests relationship-focused explanations, which show that only the $R$-$S$ join is significant (while the $R$-$T$ join plays no role). Together, these findings show the model aggregates \sql{flag} values over $S$-records linked to each $R$-record. 
Switching to \emph{Task 2}, feature-focused explanations identify \emph{no} feature column as important. How can the model predict accurately without features? Our data scientist now pivots to relationship-focused explanations and finds that the $R$-$S$ join is important. This reveals that the model relies purely on connectivity structure (i.e., counting linked records) rather than attribute values.

This workflow, where users can select explanation types tailored to databases (columns, relationships, etc.) distinguishes our framework from conventional XAI approaches \cite{breiman2001_pfi, ribeiro2016should_lime, lundberg2017unified, sundararajan2017axiomatic_IG}.
In our example, traditional feature importance methods would fail to convey that Task 2 depends solely on connectivity structure. We implemented these tasks and confirmed our method produces these explanations.
}
\end{example}

To gain insights such as those in \Cref{example:intro}, we propose a framework in which explanations are expressed as \emph{view definitions} in SQL (or any query language of choice). 
This grounds explanations in a representation familiar to database users, in contrast to common alternatives that require machine learning expertise to interpret~\cite{rauker2023inner}. For graph ML, this would require knowledge of the graph modeling details or even the inner workings of GNNs~\cite{yuan2022taxonomy}.  
 
To this aim, we deploy the classical notion of \emph{determinacy}, introduced by Nash, Segoufin, and Vianu~\cite{determinacy_2010}, that captures in a fundamental way when a collection of database views is enough to determine the result of a query. For reasons of usability, flexibility, and practical implementation, we propose a soft and statistical adaptation of this concept. 
Roughly speaking, the definition we propose identifies an explanation as \emph{accurate} if the model is likely to have \emph{similar predictions} on our database as it would have on alternative databases, as long as they agree on the views. \Cref{Fig_intro} shows a visual illustration. In addition to accuracy, we seek \emph{concise} explanations that are easy to digest. 
Our framework naturally supports the tradeoff between accuracy and conciseness, allowing users to balance interpretability with fidelity to the model's behavior. 

\rebuttal{R1D5, R2W2 and R2D}{
While our framework is \e{model-agnostic}, as it applies to any relational predictive model, efficient computation requires heuristics to avoid exhaustive search over the space of candidate explanation views, with \e{model-specific} algorithms offering further advantages by leveraging model structure. 
We instantiate the framework for the case of hetero-GNNs, for which we propose a model-specific approach for important classes of views: projections, joins over foreign keys, and selections (filters). The approach is based on learning \e{masked} variations of the GNN at hand, by applying masks on GNN components that correspond to the different view classes. 
}

Lastly, we present an empirical evaluation of our techniques on the RelBench collection of databases and tuple-level tasks~\cite{relbench}. In particular, the experiments show the tradeoff between explanation accuracy and conciseness. As a side benefit, we show that, for most RelBench tasks, we can achieve models of similar quality to the ones publicly available, while using only a fraction of the database.

\introparagraph{Our main contributions} 
\circled{\normalsize{1}} We introduce a model-agnostic framework for generating relational explanations of predictive models over relational databases, based on a statistical adaptation of determinacy. The framework uses relational queries, supports user-controlled granularity, and does not rely on internal model representations. 
\circled{\normalsize{2}} We instantiate the framework \rebuttal{R1D5, R2W2 and R2D}{for hetero-GNNs}, and develop an efficient \rebuttal{}{model-specific} approach via learnable masking that generates explanations without explicitly estimating soft determinacy. 
\circled{\normalsize{3}} We conduct extensive experiments on the RelBench benchmark across several databases and tasks.  \rebuttal{}{To facilitate comparison, we implement model-agnostic heuristic alternatives} and demonstrated that our \rebuttal{}{model-specific} method consistently produces high-quality explanations at significantly lower computational cost.

\section{Preliminaries}
\label{sec:prelim}

In this section, we define the concepts and notation that we use throughout the paper.

\subsection{Relational Database Concepts}
We begin with database concepts and terminology.

\introparagraph{Databases} A \e{relation schema} $\scr$ is associated with a sequence $\att(\scr)=(A_1,\dots,A_k)$ of distinct \e{attributes}, and a \e{key signature} $\key(\scr)=(B_1,\dots,B_\ell)$, so that $\key(\scr)$ is a subsequence of $\att(\scr)$.
Each attribute $A_i$ is associated with a (finite or infinite) \e{domain}, denoted $\dom(A_i)$. 
If $\vec C= (C_1,\dots,C_m)$ 
is a sequence of attributes of $R$, then we denote by $\dom(\vec C)$ the set $\dom(C_1)\times\dots\times\dom(C_m)$. 
A \e{tuple} $t$ over a relation schema $\scr$ is a member of $\dom(\att(R))$. 
For $i=1,\dots,k$, we denote by $t[A_i]$ the $i$-th value in $t$ (corresponding to the attribute $A_i$). 
If $\vec C=(C_1,\dots,C_m)$ is a sequence of attributes from $\att(\scr)$, then we denote by $t[\vec C]$ the tuple 
$(t[C_1],\dots,t[C_m])$. 

A \e{relation} over the relation schema $\scr$ is a finite set of tuples over $\scr$ so that $t_1[\key(\scr)]\neq t_2[\key(\scr)]$
for every two distinct tuples $t_1\neq t_2$.  If $r$ is a relation over $\scr$ and $\vec C=(C_1,\dots,C_m)$ is a sequence of attributes from $\att(\scr)$, then $r[\vec C]$ denotes the relation $\set{t[\vec C]\mid t\in r}$. We denote with $|r|$ the number of tuples in $r$.

A \e{database schema} $\scs$ is associated with a set $\rels(\scs)$ of \e{relation names}, and it maps every relation name $R\in\rels(\scs)$ to a relation schema $\scs(R)$. 
To ease the notation, we may write $\att(R)$ and $\key(R)$ as a shorthand notation for $\att(\scs(R))$ and $\key(\scs(R))$, respectively. 
In addition, a database schema $\scs$ includes a set $\fks(\scs)$ of \e{foreign-key constraints}, or \e{FK} for short, where an FK 
is an expression of the form $R[\vec C]\sqsubseteq\key(S)$, 
where $R$ and $S$ are relation names in $\rels(\scs$),
the sequence $\vec C$ consists of attributes from $\att(R)$, the sequences $\vec C$ and $\key(S)$ have the same length, and $\dom(\vec C)\subseteq\dom(\key(S))$.
A \e{database} $D$ over $\scs$ maps every relation name $R\in\rels(\scs)$ to a relation $D(R)$ over $\scs(R)$, 
so that every key constraint and every constraint in $\fks(\scs)$ is satisfied: the FK $R[\vec C]\sqsubseteq\key(S)$ is \e{satisfied} if 
$D(R)[\vec C]\subseteq D(S)[\key(S)]$.

\introparagraph{Relational queries and predictive models} 
Let $\scs$ be a database schema. A \e{query} $Q$ over $\scs$ is associated with a relation schema, which we consistently denote by $\scr_Q$, and it maps every database $D$ over $\scs$ to a relation $Q(D)$ over $\scr_Q$.

If $R\in\rels(\scs)$ is a relation name, then an \e{$R$-model} is a query 
(possibly very complicated and with learned parameters)
over $\scs$ whose purpose is to make a prediction (e.g., classification or regression) for every tuple in $R$.
Formally, an \e{$R$-model} (\e{of dimension $d$}) is a query $M$ with the following properties:
\begin{itemize}
    \item \e{existence of prediction attribute $A$}: 
    $\att(\scr_M)=(\key(R),A)$, where $A$ is an attribute with $\dom(A)=\reals^d$, where $\reals$ stands for the set of real numbers.
    \item \e{key signature agreement}: 
    $\key(\scr_M)= \key(R)$.
    \item \e{instance-prediction relation correspondence}: 
    For every database $D$, $M(D)[\key(R)] = D(R)[\key(R)]$. 
    In words, if $r=M(D)$ is the relation resulting from applying 
    $M$ to $D$ , then $r$ has the exact same key values as the $R$-relation of $D$.  
\end{itemize}
We refer to each tuple in $D(R)[\key(R)]$ 
(notice the projection on only the key attributes)
as an \e{instance} (to be classified by $M$).

Let $M$ be an $R$-model and $D$ a database, both over the schema $\scs$. 
Let $s$ be an instance (i.e., a key tuple of $D(R)$) to which we apply the model.
We denote by $M_D(s)$ the unique prediction of $M$ for $s$, that is, the value $t[A]\in\reals^d$ of the unique tuple $t\in M(D)$ with $t[\key(R)]=s$.

\introparagraph{Determinacy}
\label{par:determinacy}
We briefly recall the concept of \emph{determinacy} by \citet{determinacy_2010}. Let $\scs$ be a schema, let $V_1$, \dots, $V_m$ be queries over $\scs$, referred to as \e{views}, and let $Q$ be a query over $\scs$. We say that 
$\set{V_1,\dots,V_m}$ \e{determines} $Q$ if for all databases $D_1$ and $D_2$, if $V_i(D_1)=V_i(D_2)$ for every $i=1,\dots,m$, then $Q(D_1)=Q(D_2)$. 
Hence, knowing the result of every view suffices to determine the result of $Q$ without even looking at the database itself.

\subsection{Heterogeneous Graphs}
\label{sec:prelim_hetero-graphs}

Next, we present the concepts and notation related to machine learning over heterogeneous graphs.

\introparagraph{Hetero-graphs}
A \emph{heterogeneous featured graph}, or \emph{hetero-graph} for short, is a quadruple $\G=(V,E,\tau,\X)$ where:
\begin{itemize}
\item $V$ is a finite set of nodes;
\item $E$ is a set of \e{typed directed edges}, that is, triples $(u,\lambda,v)$ where $u$ and $v$ are nodes in $V$ and $\lambda$ is an \e{edge type};
\item $\tau(\cdot)$ is a \emph{node-type function}, mapping every node $v\in V$ to a type $\tau(v)$; and
\item $\X$ is a \e{node-feature function} that maps every node $v\in V$ to a vector $\X_v\in\reals^{d_{\tau(v)}}$ for some predefined \emph{feature dimension} $d_{\tau(v)}$ for each node type. 
\end{itemize}
Every edge type $\lambda$ in $G$ is associated with a \e{source type} $\src(\lambda)$ and a \e{target type} $\tgt(\lambda)$, and we require all edges $e=(u,\lambda,v)$ to be consistent in that $\tau(u)=\src(\lambda)$ and $\tau(v)=\tgt(\lambda)$.
We denote by $\tau(V)$ the set $\set{\tau(v)\mid v\in V}$. 
If $\alpha\in\tau(V)$, then we denote by $V[\alpha]$ the set of nodes $v\in V$ with $\tau(v)=\alpha$.

We denote by $\N^\vin(v)$ (resp., $\N^\vout(v)$)
the set of incoming (resp., outgoing) neighbors of $v$, that is, the set $\set{u\mid \exists\lambda[(u,\lambda,v)\in E]}$
(resp., $\set{u\mid\exists\lambda[ (v,\lambda,u)\in E]}$). 
If $\lambda$ is an edge type, then we denote by $\N^\vin_\lambda(v)$ the set of nodes $u\in\N^\vin(v)$ such that $\tau(u,\lambda,v)\in E$; analogously, we denote by $\N^\vout_\lambda(v)$ the set of nodes $u\in\N^\vout(v)$ with $\tau(v,\lambda,u)\in E$. 

\introparagraph{Node-centric models}
Let $\G=(V,E,\tau,\X)$ be a hetero-graph, and let $\alpha\in\tau(V)$ be a type, which we refer to as a \emph{prediction entity type}. A \emph{$\alpha$-model} (of dimension $d$) 
is a function $M: V[\alpha]\rightarrow \reals^d$, mapping every $\alpha$-node $v$ to a predicted vector $M(v)$. Note that the dimension $d$ depends on the task (e.g., it is one in binary classification and regression, and $k-1$ in $k$-class classification), and is not necessarily the same as $d_\alpha$.

\subsection{Heterogeneous Graph Neural Networks}
\label{sec:prelim_hetero-gnn}

A GNN defines a model over a graph by applying message passing to produce a vector representation of each node; in each step, called \e{layer}, the vector of a node is updated by aggregating the vectors of its neighbors, referred to as \e{messages}. 
A \e{hetero-GNN} operates over a hetero-graph similarly, except that messages of each edge type are aggregated separately. We give a more precise definition of a typical hetero-GNN next. We note that our definition is slightly more general and detailed than common definitions (e.g.,~\cite{barcelo2022weisfeiler_relational}), stressing the role of labels and directionality in the GNN.

Let $\G=(V,E,\tau,\X)$ be a hetero-graph and $\alpha\in\tau(V)$. 
A hetero-GNN with $L$ layers defines an $\alpha$-model using the functions $\vec h^{(\ell)}$ 
and $\A^{(\ell)}$, for $\ell=0,\dots,L$, constructed inductively as follows. 
First, define $\vec h^{0}(v)=\X_v$ for all $v\in V$. For $\ell=0,\dots,L-1$, define
$$
\A^{(\ell)}(v,\lambda,o)=\agg(\bag{\vec h^{(\ell)}(u) \mid u\in \N^o_\lambda(v)})
$$ 
for every node $v$, edge label $\lambda$ and orientation $o\in\set{\vin,\vout}$ 
such that $\N^o_\lambda(v)$ is defined. 
Here, $\agg$ is an aggregation function such as average or max, and $\bag{\cdot}$ is the bag (multiset) notation.
Next, for $\ell>0$ we define
\begin{equation}
\label{eq:gnn-upd}
\begin{split}
\vec h^{(\ell)}(v) = \upd(
&\vec h^{(\ell-1)}(v), \\
&\A^{(\ell-1)}(v,\lambda_1,o_1),\dots, \\
&\A^{(\ell-1)}(v,\lambda_q,o_q))
\end{split}
\end{equation}

for some ordering $(\lambda_1,o_1),\dots,(\lambda_q,o_q)$ of all pairs $(\lambda,o)$ where $\N^o_\lambda(v)$ is defined. Here, $\upd$ is an \e{update function}, which is realized as a neural model. A typical example of model architecture for hetero-GNNs \cite{schlichtkrull2018-RGCN} involves a linear combination followed by an activation function (e.g., ReLU).
The model $M$ is given by $M(v) = \mlp(\vec h^{(L)}(v))$ %
where $\mlp$ is a multilayer-perception function that transforms the last layer into the actual prediction. 
The learnable parameters are incorporated in the 
functions $\agg$,
$\upd$, and $\mathrm{MLP}$.

\subsection{Relational Learning via Heterogeneous Graphs}
\label{sec:prelim_rdl}

A common way of applying deep learning over relational databases is to convert the database $D$ into a hetero-graph $\G_D$, and then learn and apply a hetero-GNN over $\G_D$~\cite{cvitkovic2020supervised, cappuzzo2024relational, fey2024position, 4dbinfer_wang2024}. Next, we describe this process more precisely for a typical conversion, known as the \emph{relational graph} due to the \textsf{r2n} (i.e., row-to-node) method~\cite{4dbinfer_wang2024}. 

Let $\scs$ be a schema, and let $D$ be a database over $\scs$. 
The graph $\G_D=(V,E,\tau,\X)$ is defined as follows. The set $V$ of nodes consists of the pairs $(R,t)$, where $R$ is a relation name and $t$ is tuple in $D(R)$:
$$V=\set{(R,t)\mid R\in\rels(\scs)\land t\in D(R)}\,.$$ 
For a node $v=(R,t)$, we define $\tau(v)=R$, that is, the type of a node is the name of its relation.
There is a directed edge $e=(v,\varphi,u)$ from a node $v=(R,t)$ to a node $s=(S,t')$ whenever the two are connected by a reference via an FK $\varphi$ %
of the form $R[\vec C]\sqsubseteq \key(S)$; in particular, the type of the edge is the FK that it corresponds to. 
Finally, the feature vector $\X$ is defined as follows. Let $R$ be a relation name with
$\att(R)=(A_1,\dots,A_k)$. 
For each attribute $A$ we assume a trainable encoder 
$\enc_{R.A}:\dom(A)\rightarrow\reals^{d_{R.A}}$ 
for some predefined dimension $d_{R.A}$. 
In addition, for each relation name $R$ we assume a trainable encoder 
$\enc_{R}:\reals^{d_{R_{\mathit{enc}}}}\rightarrow\reals^{d_{R}}$, 
for some dimension $d_{R}$ and $d_{R_{\mathit{enc}}} = \sum_{i=1}^k d_{R.A_i}$. For a node $v=(R,t)$ we then
define:
\begin{equation}
\label{eq:encoder}
\X_v=\enc_{R} \left( \enc_{R.A_1}(t[A_1])\oplus
\cdots\oplus
\mathrm{Enc}_{R.A_k}(t[A_k]) \right)
\end{equation}
where $\oplus$ stands for vector concatenation. 
Notice that this formulation allows us to exclude some attributes $A$ of a relation $R$ from the encoding, simply by setting  $\mathrm{Enc}_{R.A_k}$ to map every value to the empty vector. For example, in our implementation, we exclude key and foreign-key attributes from the features of a record.

Finally, let $R$ be a relation name in $\rels(\scs)$. An $R$-model $M'$ over $\G_D$ defines an $R$-model $M$ over the database $D$ in a straightforward way: For a tuple $t\in D(R)$, we define 
$$
M(t[\key(R)]) = M'(v)
$$
where $v$ is the node $(R,t)$ in the node set $V[R]$.

\section{RDL Explanations via Soft Determinacy}
\label{sec:framework}

In this section, we introduce our framework, where we leverage the concept of determinacy to produce explanations.

\subsection{Explanation Views}

Let $\scs$ be a schema with a relation name $R$, and let $M$ be an $R$-model. Let $D$ be a database over $\scs$. We think of $M$ as a black-box complex model, involving a considerable number of computational components and learned parameters. 
By an \e{explanation} of $M$'s behavior on $D$, we refer to a collection of views $V$ (called \e{explanation views}), each indicating a focused component of $D$ that is important for $M$. 
Next, we define the explanation views more formally.

We assume an \e{explanation language} $\exlang$, which is simply a query language that we use as a formalism for defining explanation views. 
An explanation view $V$ states what component of the database $D$ (existing or derived) is relevant for $M$ to make the prediction $M_D(s)$ on any instance $s$; for that, we assume that $V$ is parameterized by $s$, denoted $V\angs{s}$. 
This means that $V\angs{s}$ is an ordinary database query with the result $V\angs{s}(D)$, for every instance $s$. 
Recall that $s$ is represented as a key tuple in $D(R)$; 
hence, to represent explanation views, we assume that queries in $\exlang$ can refer to the attributes of $\key(R)$ as constants. 

\begin{example}
\label{ex:view1}
As an example, suppose that our schema consists of $R(A)$ and $T(A,B)$, and $s \in D(R)$. The following explanation view (in SQL) returns the $T$-tuples that connect to $s$:
\begin{align*}
\sql{create} \sql{ view $V\angs{s}$ as } \sql{select} \sql{ $*$ from $T$ } \sql{where}  \sql{ $T.A=s.A$ }
\end{align*}
\end{example}

Note that the parameterized view $V$ does not necessarily have to refer to its parameter in its definition. 
If $V$ ignores its parameter $s$, we say that it is \e{instance-agnostic}, otherwise it is \e{instance-specific}. 
Formally, $V$ is instance-agnostic if, for each database $D$ and instances $s$ and $s'$ it holds that $V\angs{s}(D)=V\angs{s'}(D)$. 
As an example, the following explanation view is instance-agnostic, as it refers to all tuples of $T$, regardless of their relationship to $s$: \textsf{select} \textsf{ $*$ from $T$}.%
Finally, by an \emph{explanation} $E$ we mean a finite subset of concrete queries from $\exlang$. The quality of an explanation is measured using two criteria: \e{determinacy} (i.e., whether $E$ is enough to determine the behavior of $M$) and \e{conciseness} (i.e., how easy it is to grasp $E$).

\subsection{Soft Determinacy}

Consider an explanation $E$ of an $R$-model $M$ on a database $D$. Intuitively, $E$ ``sufficiently determines'' $M$ if $M$ behaves similarly when $D$ changes, as long as the result of $E$ is not affected by this change. 
Ideally, for any two databases that agree on the explanation views in $E$, the output of 
$M$ should be identical. Viewing $M$ as a query, this strict sufficiency notion is precisely \e{determinacy}, as described in \Cref{par:determinacy}. To capture the imperfect nature of explanations
, we relax the notion of determinacy in two dimensions. First, we do not require exact determinacy;
instead, we account for \e{approximate agreement} by quantifying the distance between model predictions before and after the change (see \cref{Fig_intro}).
Second, we move away from deterministic determinacy via 
a probabilistic adaptation. For that, we require determinacy \e{in expectation}, using our original database $D$ as reference and assuming a probability distribution of \e{contingency databases}. 
The two relaxations lead to an approximate probabilistic notion of determinacy, namely \e{soft determinacy}. 

Formally, we assume a distribution $\Delta$ over contingency databases that agree on the schema $\scs$ of $D$. 
To be comparable to $M(D)$, we require each database $D'$ in this distribution to agree with $D$ on the keys of $R$, that is: $[D'(R)[\key(R)]=D(R)[\key(R)]\,$.

To define approximate agreement, suppose that $M$ maps an instance $s$ to $a$, and an instance $s'$ to $a'$. 
We use a distance metric $\dist$ that quantifies how far $a$ is from $a'$ as $\dist(a,a')$. For example, consider the special case where $M$ is a binary classifier with predictions $\in \{0, 1\}$; then, it makes sense to use $d$ that checks for identity: 
$
\dist(a,a')$ is
0 if $a=a'$, and
1 otherwise.

Using the above, we arrive at the soft determinacy property for an explanation $E$, quantified by \e{deviation from determinacy}, so that lower deviation implies a better explanation. 

\begin{definition}[Deviation from Determinacy]
\label{def:soft_determinacy}
For an instance $s$, and a distribution $\Delta$ of contingency databases, 
the (instance-specific) deviation from determinacy $\devi_\Delta(E, s)$ of an explanation $E$ is the \e{expected distance} between the prediction for $s$ on $D$ and the prediction on a contingency $D'$, given that $D'$ respects the views: 
\begin{align*}
    \devi_\Delta(E, s)\defeq
    \underset{D'\sim\Delta}{\mathbb{E}}
    \Big[& \dist(M_D(s),M_{D'}(s))
    \;\Big| \bigwedge_{V\in E}V\angs{s}(D)=V\angs{s}(D')
    \Big]
\end{align*}   
Deviation from determinacy of $E$ is the empirical mean over instances: $\devi_\Delta(E)\defeq \frac{1}{|D(R)|}\sum_{s\in D(R)[\key(R)]} \devi_\Delta(E, s)$.
\end{definition}

\subsection{Explanation Conciseness}
\label{sec:exp_objective}

We associate with each $V\in\exlang$ a \e{cost} related to its complexity. 
This cost can be syntactic, e.g., the number of symbols involved in the SQL representation. 
It can also be semantic, e.g., the number of tuples in the view $|V(D)|$. 
We denote this cost by $\cost(V)$. We then define the cost of an explanation $E$ as $\cost(E)=\sum_{V\in E}\cost(V)$. %

The computational problem of finding the best explanation can be phrased as an optimization problem:

\begin{definition}[Soft determinacy explanations]
\label{def:exp_objective}
Given an $R$-model $M$ and a database $D$, the soft determinacy explanation problem asks for a solution $E\subseteq\exlang$ to the following optimization objective:
\begin{align*}
\min_{E} [\devi_\Delta(E) + \lambda \cdot \cost(E)], 
\end{align*}
for a trade-off captured by $\lambda \in \reals, \lambda \geq 0$.
\end{definition}    

Clearly, there is a natural trade-off between achieving high soft determinacy (minimization of $\devi_\Delta(E)$) and low explanation complexity (minimization of $\cost(E)$). Consider the extreme case of $E$ containing the entire schema $\scs$. In this case, $E$ fully determines the model $M$, but also does not reduce the complexity of $M$ or $\scs$. 
On the other hand, 
a void explanation $E=\emptyset$ has minimal complexity, but likely does not capture any insight on the model's behavior.

\rebuttal{R3C3 and R3D3}{
The optimization problem in \Cref{def:exp_objective} displays the challenges commonly associated with bilevel optimization \cite{dempe2002bilevel, colson2007bilevel}. It involves two levels: the outer selects explanation views to minimize cost, while the inner estimates $\devi_\Delta(E)$ by sampling and evaluating the model on multiple perturbed databases that respect the constraints imposed by the outer explanation selection. The outer level, i.e., searching among exponentially-many subsets of database components to find those that best preserve the model behavior, requires exploring a vast combinatorial space (see \Cref{ex:hardness}).
}
\begin{example}
\label{ex:hardness}
\rebuttal{}{
Consider the case where we seek explanation views that only involve projections over a single table with binary attributes, and we require exact determinacy ($\devi_\Delta(E) = 0$). Suppose our model $M$ is a disjunction (\sql{OR}) over all attributes, i.e., $M$ predicts $1$ for all tuples except the all-zeros tuple $(0,\ldots,0)$. A minimal explanation corresponds to a minimal set of columns such that every tuple has at least one $1$ in those columns. This is precisely the classical hitting set problem, which is NP-hard. This example demonstrates that even for the simplest explanation language and model structure, finding optimal explanations requires considering exponentially many column subsets. The problem becomes more complex with more expressive explanations not restricted to projections.
}
\end{example}

To address these challenges, in \Cref{sec:framework_instantiation} we discuss several choices for realizing this objective into a more practical optimization problem. Additionally, in \Cref{sec:model_specific} we propose a learning-based solution for efficient explanation view discovery for hetero-GNNs.

\section{Instantiating the Framework for Selected SQL Fragments} 
\label{sec:framework_instantiation}

The framework presented in \Cref{sec:framework} is abstract and requires the choice of an explanation language $\exlang$ together with a concrete metric for conciseness and the distribution of contingency databases $\Delta$.
In this section, we give specific instantiations that we deem useful and general. In the next sections, we propose heuristic implementations of these over GNN models and test them in experiments.

\subsection{Explanation Languages ($\exlang$'s)}
\label{sec:exlang}

Our explanation framework highlights the database components that provide the most useful information for the model. These may include important features, important feature combinations, or influential values and value ranges of particular features. In SQL terms, these explanations can be expressed through three core operations: projecting only the relevant attributes of relations, joining relations to reveal informative combinations, and selecting tuples that satisfy specific conditions. To capture these explanations, we focus on three simple SQL fragments utilizing the fundamental relational algebra operations: projections, joins, and selections. Throughout the remainder of this paper, we restrict our attention to instance-agnostic explanation views, for which we define suitable choices of $\exlang$ and later propose efficient algorithms. We leave the study on instance-specific explanations for future work.

To define each $\exlang$ precisely, we classify the attributes of a relation schema $\scr$ in a database schema $\scs$ into three categories. Given $\att(\scr) = (A_1, \dots, A_n)$, an attribute $A_i$ is a \e{key attribute} if it appears in the key signature $\key(\scr)$. It is a \e{foreign-key attribute} if it appears on the left-hand side of a foreign-key constraint in $\fks(\scs)$. Any attribute that is neither a key nor a foreign key is called a \e{data attribute}. In the definitions of $\exlang$ that follow, these three attribute types are treated separately.

(1) The language $\proj$ can be used to highlight subsets of data attributes corresponding to important features, while always preserving all key attributes and foreign-key attributes. Formally, $\proj$ contains views of the form:
\begin{align*}
&\textsf{select} \textsf{ $\key(R)$, $\vec F$, $A_1,\dots,A_k$ from $R$}
\end{align*} 
which project onto a chosen subset of data attributes $A_1, \dots, A_k$, while retaining the sequence $\key(R)$ of key attributes and the sequence $\vec F$ of foreign-key attributes of $\scs(R)$. %

(2) The language $\join$ consists of foreign-key joins. Each explanation view in $\join$ is associated with a foreign-key constraint $R[\vec C] \sqsubseteq \key(S)$ in the schema, and has the form:
\begin{align*}
&\textsf{select} \textsf{ $*$ from $R$ as $r$, $S$ as $s$ where $r[\vec C] = s[\key(S)]$}
\end{align*}  
These views retain all tuples produced by joining $R$ and $S$ using foreign-key references. Note that more complex join patterns can be captured by a set of multiple views. Because $\join$ focuses on foreign keys, it always preserves all key attributes and data attributes of relations that appear in at least one explanation view.

(3) The language $\select$ focuses on filtering tuples based on attribute values. As with $\proj$, the emphasis is on data attributes, while key and foreign-key attributes are always preserved. Formally, $\select$ contains views of the form:
\begin{align*}
&\textsf{select}  \textsf{ $*$ from $R$ where $\phi$}
\end{align*} 
where $\phi$ is a selection predicate that identifies an informative subset of tuples. 
For example, consider a database with a \sql{Product} $P$ relation. The explanation view: ``\sql{select $*$ from $P$ where $P.price<50$ and $P.cat=$ Electronics}'' highlights the group of affordable electronics. %

\introparagraph{Cost model} 
Explanation conciseness requires a precise definition of the cost model so that every explanation view $V \in E$ is associated with a value $\cost(V)$. Here we propose a simple cost model that focuses on minimizing the description cost of explanation views, making them easier for a user to understand. For $\proj$, $\cost(V)$ is the number of highlighted data attributes, i.e., $\cost(V) = k$. For $\join$, each binary join incurs a cost of $1$, i.e., $\cost(V) = 1$. For $\select$, $\cost(V)$ depends on the complexity of the selection predicate $\phi$. Later, we will consider disjunctions of the form \textsf{$\phi_1$ or $\dots$ or $\phi_l$}, where each $\phi_i$ is an atomic predicate. In this case we simply assign $\cost(V) = l$, i.e., the number of atomic predicates involved.

\introparagraph{Language composition} We consider combinations of the three languages by constructing composite explanation views in the relational algebra sense. Importantly, these combination languages produce composite views rather than unions of separate explanation views belonging to one of the three defined languages. For composite views, the cost model is naturally extended: we sum the number of data attributes in projections, the number of joins, and the number of selection predicates involved.

\subsection{Database Perturbations}
\label{sec:perturbations}

Recall that the definition of deviation from determinacy (\cref{def:soft_determinacy}) allows the liberty to choose any distribution $\Delta$ of contingency databases, as long as they agree with $D$ on the explanation views $E$. 

Let us consider two extreme scenarios to illustrate the impact of $\Delta$. First, consider a $\Delta$ over databases $D'$ that differ from $D$ in just one value of a single tuple. 
A well-behaved model would likely give a similar output on $D$ and $D'$, irrespective of the chosen explanation $E$. Therefore, such $\Delta$ would not yield a discriminative criterion between candidate explanations. On the other hand, if $\Delta$ is the uniform distribution over all possible $D'$, no matter how different they are from $D$, it is not clear how one could obtain samples from $\Delta$. More importantly, the sampled $D'$ could be unrepresentative for the application domain of the original database $D$. 
Ideally, we would like to limit the explanatory power of $E$ to the useful regime, i.e., restricting $\Delta$ to \emph{a local neighborhood} of $D$ that captures realistic variations while preserving relevance to the application domain.

\rebuttal{R1D3, R2W1 and R2D1}{
Defining input perturbations is a fundamental challenge across ML explanation approaches that replace insignificant features with uninformative values. 
\emph{Permutation Feature Importance} (PFI), a well-known model-agnostic feature selection technique \cite{breiman2001_pfi, molnar2020interpretable}, replaces feature values by permuting them across instances, preserving marginal statistics while breaking correlations with the target. However, this approach may generate unrealistic data variations that violate feature correlations, thus forcing models to extrapolate \cite{fisher2019_pfi, hooker2021unrestricted_permutations}. PFI variants address this by sampling from the conditional distribution $P(x_j | x_{-j})$ of the perturbed feature $x_j$ \cite{zintgraf2017visualizing_conditional, fisher2019_pfi, debeer2020cpi}, but become computationally inefficient without strong assumptions like feature locality \cite{zintgraf2017visualizing_conditional}, assumptions more natural for images than tabular data. Subgroup-based methods partition instances into groups in order to reduce dependence before permuting \cite{molnar2024conditional_subgroups}, however they introduce additional complexity by training auxiliary models (e.g., decision trees) to identify appropriate subgroups for each feature. 

We adapt these insights to relational databases by designing language-specific perturbations. 
For each explanation language $\exlang$, we provide strategies for realizing $\Delta$ by perturbing the original database $D$. Each strategy neutralizes information outside the explanation while respecting the explanation views. Following the PFI literature's concern with correlation preservation \cite{fisher2019_pfi, hooker2021unrestricted_permutations, molnar2024conditional_subgroups}, we test both correlation-preserving and correlation-breaking variants in \Cref{sec:experiments} and find results are mostly robust to this choice}.

\introparagraph{$\proj$} 
We permute columns not appearing in the explanation (independently or jointly), inspired by PFI. 

\begin{figure}
    \centering
    \includegraphics[scale=0.31]{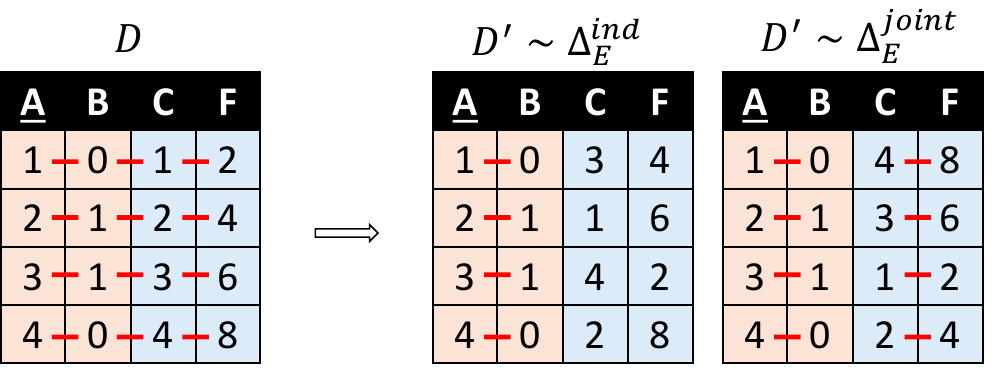}
    \includegraphics[scale=0.31]{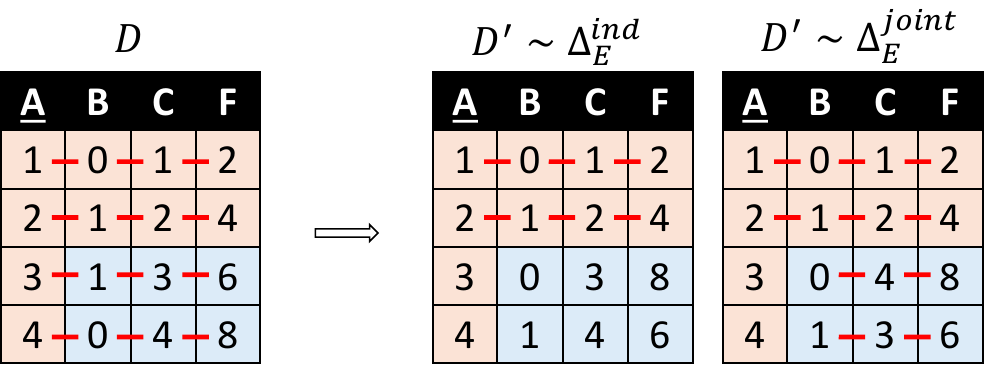}
    \caption{Example permutations for $\proj$ (upper) and $\select$ (lower). 
    For $\proj$, only attributes $A$ and $B$ (in orange) are in $\attrs(E)$
    of explanation $E$.
    For $\select$, the tuples (in orange) $\{(1,0,1,2), (2,1,2,4)\}$ are in $\tups(E)$. The key attribute $A$ is always retained.
    }
    \label{fig:column_perturbations}
\end{figure}

\begin{example}
\label{example:join permutation}
The example in \Cref{fig:column_perturbations} shows two column-wise permutation strategies. 
The attributes $A$ and $B$ are part of the explanation $E$, whereas $C$ and $F$ are not. 
Thus, we permute $C$ and $F$ to produce different $D'$. 
We can either apply the permutations to both attributes $C$ and $F$ jointly (left) or independently (right). 
Note that the latter completely breaks correlations between $C$ and $F$, possibly creating unnatural tuples in $D'$. In this example, we have $F = 2 \cdot C$ for all tuples in $D$. Joint permutation preserves this property in $D'$.
\end{example}

Formally, let $\attrs(E)$ be the set of attributes used in a projection explanation $E$. Recall that $\attrs(E)$ includes all key attributes. Because explanation views must be respected by $\Delta$, we only permute attributes not in $\attrs(E)$. 
We define two permutation distributions that differ in how they handle correlations among non important attributes. 
The \emph{joint column-wise permutation distribution} $\Delta^{\mathsf{joint}}_{E}$ is the uniform distribution over all databases $D'$ where, for each relation in $D$, attributes $\notin \attrs(E)$ are permuted together by the same permutation. \rebuttal{}{This preserves correlations among unimportant attributes while destroying correlations between important and unimportant attributes, and between the target and unimportant attributes.}
The \emph{independent column-wise permutation} distribution $\Delta^{\mathsf{ind}}_{E}$ is the uniform distribution over all databases $D'$ where each attribute $\notin \attrs(E)$ is permuted independently. \rebuttal{}{This additionally breaks correlations among unimportant attributes.}

\introparagraph{$\join$}
The goal of a perturbation is to make joins uninformative unless they appear in $E$. Since the joins we consider always correspond foreign-key constraints, we apply a perturbation on the foreign key side. This perturbs the target join without affecting other joins on the side of the primary key. 

\begin{figure}[t]
    \centering
    \includegraphics[scale=0.31]{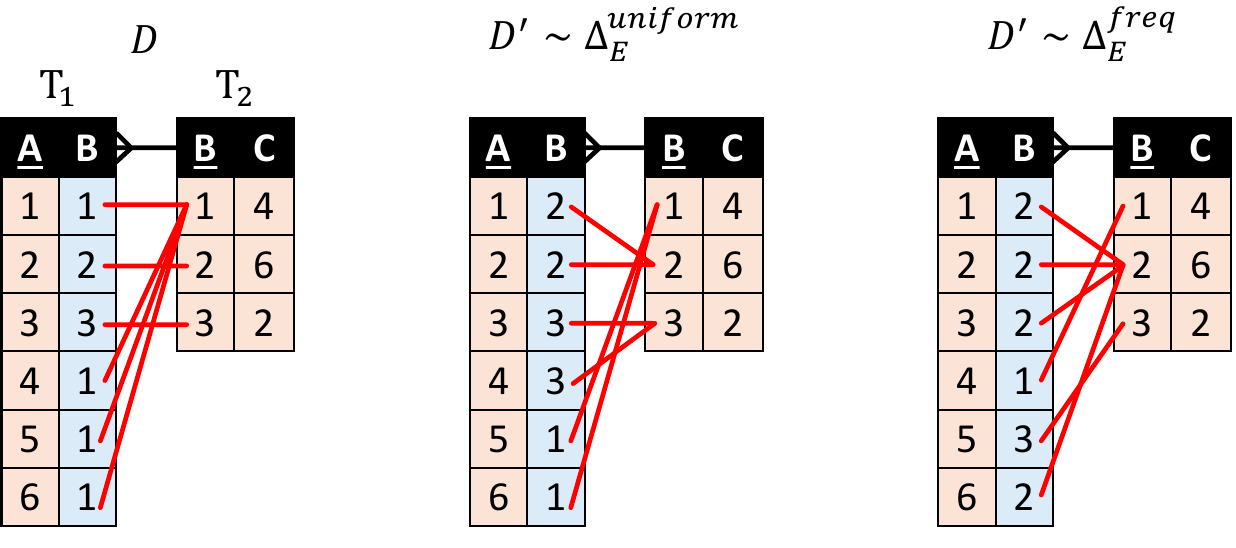}
    \caption{\Cref{example:FKjoin}: foreign key perturbations for $\join$. 
    Here, the foreign key $T_1.B$ is perturbed as it is not in $\fkeys(E)$.}
    \label{fig:fk_perturbations}
\end{figure}

\begin{example}
\label{example:FKjoin}
\Cref{fig:fk_perturbations} (left) shows perturbations applied to the primary-foreign pair $T_2.B$ - $T_1.B$. 
One could consider permuting the foreign key $T_1.B$. But this strategy would not be effective in breaking the predictive signal encoded in the actual join condition as it would retain the number of primary-foreign key pairs per unique value of $T_1.B$, i.e., $1$ would appear 4 times while $2$ and $3$ would each appear once. In order to break the join information, we randomly replace each value of $T_1.B$ with a value in the domain $\{1,2,3\}$, either uniformly at random (right) or mimicking the underlying frequency distribution of foreign keys (middle). 
In the former each value in $\{1,2,3\}$ appears $2$ times while in the latter $2$ appears 4 times while $1$ and $3$ each appear once, preserving the frequencies $4, 1, 1$. 
\end{example}

We consider two alternative strategies: In \emph{uniform replacement} $\Delta^{\mathsf{uniform}}_E$, each foreign key value is replaced independently and uniformly at random from its domain. In the \emph{frequency-preserving replacement} $\Delta^{\mathsf{freq}}_E$, the frequences of foreign key values in the original relation $r$ and the perturbed one $r'$ match: if a value occurs $i$ times in $r$, then there exists some value in $r'$ that occurs $i$ times, although not necessarily in the same tuples. \Cref{fig:fk_perturbations} shows an example, with $\fkeys(E)$ denoting the foreign keys in explanation $E$.

\introparagraph{$\select$} 
We adapt the column-wise permutation strategies used in the case of $\proj$ for \e{all} data attributes, 
but applied only to the tuples that do not satisfy the selection condition of the explanation view. \Cref{fig:column_perturbations} (lower) shows an example, with $\tups(E)$ denoting the set of tuples that satisfy the selection condition.

\section{GNN-Specific Approach}
\label{sec:model_specific}

\rebuttal{R1D5, R2W2 and R2D}{
We now instantiate our framework (\Cref{sec:framework}) for the case of GNNs. We present a model-specific approach that leverages properties of the model to efficiently solve the optimization in \Cref{def:exp_objective}: differentiability enables explanation discovery via gradient-based learning, while the model structure (i.e., feature encoding followed by message-passing) provides natural intervention points.
}

\subsection{Mask Learning: From Discrete to Continuous Optimization}
\label{sec:model_specific_justification}
\rebuttal{R1W1 and R1D9a, R3D4}{
As discussed in \Cref{sec:exp_objective}, a direct solution to \Cref{def:exp_objective} is not practically feasible. We now show how mask learning naturally arises as a continuous relaxation of the discrete soft determinacy problem, focusing on $\proj$ explanations.

\introparagraph{Discrete Selection Problem}
Recall that soft determinacy (\Cref{def:soft_determinacy}) measures sufficiency of an explanation $E$ by evaluating the expected distance between $M_D(s)$ and $M_{D'}(s)$ over contingency databases $D'$ that respect the explanation views. For $\proj$ explanations over $N$ data attributes, we seek a binary selection vector $\vec a \in \{0,1\}^N$ indicating which attributes to include in the explanation. To evaluate soft determinacy, we construct contingency databases by applying the joint column-wise permutation: for each tuple $t$, we replace the value of each attribute $A_i$ as:
$$
t[A_i] \leftarrow a_i \cdot t[A_i] + (1 - a_i) \cdot t'[A_i]
$$
where $t'$ is a randomly selected tuple from the same relation, and $a_i \in \{0,1\}$ determines whether attribute $A_i$ is fixed ($a_i=1$) or replaced ($a_i=0$). The soft determinacy objective then becomes:
$$
\min_{\vec a \in \{0,1\}^N} \quad \mathbb{E}_{s} \, \mathbb{E}_{\omega} \left[ 
\dist(M_D(s), M_{\vec a, \omega, D}(s)) \right] + \lambda ||\vec a||_0
$$
where $s$ is an instance, $\omega$ represents the randomness in selecting replacement tuples $t'$, and $||\vec a||_0$ counts the number of selected attributes. However, optimizing over $\vec a \in \{0,1\}^N$ requires searching through $2^N$ combinations, which is intractable.

\introparagraph{Continuous Relaxation via Direct Masking}
A natural approach is to relax the binary constraint: replace $\vec a \in \{0,1\}^N$ with continuous mask values $\vec m \in [0,1]^N$. This yields the continuous optimization problem:
$$
\min_{\vec m \in [0,1]^N} \quad \mathbb{E}_{s} \, \mathbb{E}_{\omega} \left[ 
\text{loss}(M_D(s), M_{\vec m, \omega, D}(s)) \right] + \lambda ||\vec m||_1
$$
where we apply the masking operation at the feature encoding level. Concretely, let $\vec x \in \reals^d$ be a feature vector for attribute $A_i$. Applying mask component $m_i \in [0, 1]$ means replacing it with:
$$
\mu_\omega(\vec x, m_i) = m_i \cdot \vec x + (1 - m_i) \cdot \vec u_\omega,
$$
where $\vec u_\omega$ is the encoding of the replacement value from tuple $t'$, i.e., $\vec u_\omega = \enc_{T.A_i}(t'[A_i])$. Note that the $\ell_1$ penalty $||\vec m||_1$ serves as a convex relaxation of the $\ell_0$ cardinality constraint, encouraging sparsity \cite{hastie2015statistical_wainwright}. Additionally, the function $\text{loss}()$ is now a \emph{continuous distance metric} (e.g., cross entropy, mean squared error).

This continuous relaxation enables gradient-based optimization using automatic differentiation frameworks. When $m_i=0$, attribute $A_i$ is completely replaced, and as $m_i$ increases toward 1, the original value is progressively restored. In practice, we use stochastic optimization \cite{KingmaB15} as follows: for each optimization step we first sample a mini-batch of instances from $D(R)$ to compute the empirical loss, and then we apply one realization of the random replacement $\omega$ (e.g., one random permutation of tuples in each relation of $D$). After optimization, we threshold masks using $m_i \geq \delta$ to recover discrete explanations, a common technique for mask discretization \cite{amara2022graphframex}. 

\introparagraph{Alternative Approaches}
While direct masking has proven effective for interpretability in images \cite{fong2017masking} and graph neural networks \cite{ying2019gnnexplainer}, other techniques also exist for continuous relaxation of discrete selection problems. Policy gradient methods such as REINFORCE \cite{williams1992} treat selection as a stochastic policy, computing gradients via the likelihood ratio trick. Reparameterization approaches like Gumbel-Softmax \cite{jang2017,maddison2014,maddison2016} introduce auxiliary random variables to create differentiable approximations of discrete sampling. For example, L2X \cite{chen2018learning} uses these techniques to compute local explanations via feature selection with a fixed budget. 

We adopt direct masking due to several practical advantages. First, it provides computational efficiency: requiring only a single forward pass per mask-learning step without sampling auxiliary variables, and producing stable gradients at all points in the continuous domain $[0,1]$. Second, it naturally fits for the complex RDL model, where computation involves multiple stages (feature encoding, multi-layer
message passing), allowing unified intervention across different explanation languages $\exlang$ by modulating continuous internal representations. The latter enables joint optimization for composite languages, as the same masking operation applies regardless of whether we mask attributes ($\proj$), foreign keys ($\join$), or tuple groups ($\select$).
}

\rebuttal{}{
\introparagraph{Quality of Mask Learning Solutions}
We discussed how mask learning arises as a natural continuous relaxation of the discrete soft determinacy problem. Specifically, for $\proj$ and discrete masks ($\vec m \in \{0,1\}^n$), the objectives are equivalent: $\devi_{\Delta^{\mathsf{joint}}_{E}}(E) = \mathbb{E}_{s} \, \mathbb{E}_{\omega} \left[\text{loss}(M_D(s), M_{\vec m, \omega, D}(s)) \right]$. 
In practice, 
we learn continuous masks and threshold to obtain discrete explanations. We now argue that this approach yields high-quality solutions.

Consider the simplified setting of a binary classifier $M_{\mathcal{C}}$ that relies on \emph{precisely} $k$ data attributes.
First, there exists a high-quality $\proj$ explanation $E^*$ for $M_{\mathcal{C}}$: since $M_{\mathcal{C}}$ uses exactly $k$ columns, the $k$-column explanation selecting these achieves cost $\ell^* = \lambda \cdot k$ (zero task loss plus regularization). 
Second, the mask learning objective promotes this solution: any \emph{smaller} mask (with $k' < k$ attributes) will incur substantial task loss since important features are masked, thus predictions flip with probability $\epsilon>0$, yielding expected cost $\ell' = \epsilon + \lambda \cdot k' >> \ell^*$ for appropriate $\lambda$. Conversely, any \emph{larger} mask (with $k'' > k$ attributes) incurs higher regularization cost $\lambda \cdot k'' > \ell^*$ without reducing task loss. 
Hence, the mask learning objective promotes the optimal $k$-column explanation $E^*$.
}

\subsection{Language-Specific Masking}

\begin{figure}
    \centering
    \def\keyA{$\underline{A}$}
    \def\keyD{$\underline{D}$}
    \def\en#1{\mathrm{Enc}_{#1}}
    \scalebox{0.8}{
    \colorlet{red}{black!30}
    \input{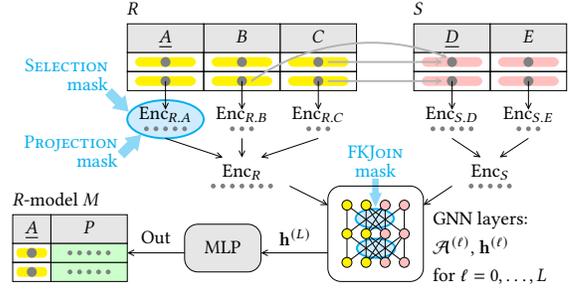}
    \definecolor{red}{rgb}{1,0,0}
    }
    \caption{RDL model pipeline with hetero-GNNs. The figure shows where different masks are located in the pipeline.
    }
    \label{fig:rdl_model}
\end{figure}

Depending on $\exlang$, masking is applied at the feature encoding stage or the message-passing stage of the model, as depicted in \Cref{fig:rdl_model}.

\introparagraph{(1) $\proj$}
We apply mask components attribute-wise in the feature encoding space. Since $\proj$ necessarily includes all primary and foreign keys, we restrict masks to data attributes.

For each relation $R \in \scs$, we assign one mask component $m_{R.A_i} \in [0,1]$ per data attribute $A_i$, $i=1, \dots, n$. For each node $v = (R,t)$, the same mask vector $(m_{R.A_1}, \ldots, m_{R.A_n})$ is applied at feature encoding, changing \Cref{eq:encoder} as follows:
\begin{equation}
\begin{split}
\X_v=\enc_{R} ( 
&\mu(\enc_{R.A_1}(t[A_1]), m_{R.A_1})\oplus\cdots \oplus\\ 
&\mu(\mathrm{Enc}_{R.A_n}(t[A_n]), m_{R.A_n}) )
\end{split}
\notag
\end{equation}
Recall that $\mu(\vec x, m)$ applies the mask component $m$ to the encoding vector $\vec x$. Every time we apply masking attribute-wise, we first consider a random permutation to the tuples of $D(R)$, which assigns for each tuple $t$ a replacement tuple $t'$. Then to apply masking to $v = (R,t)$ for each attribute $A_i$ we use the replacement vector $\vec u_i = \enc_{R.A_i}(t'[A_i])$, i.e., the encoding of the value of $A_i$ in $t'$. This ensures the replacement of the $t[A_i]$ encoding is uninformative, yet realistic. Masks for all relations and attributes are learned jointly.

\introparagraph{(2) $\join$}
We apply masks at message-passing, learning a mask value $m_{\lambda_i} \in [0,1]$ for every edge type $\lambda_i$. Since every foreign-key pair corresponds to an edge type (see~\Cref{sec:prelim_rdl}), $m_{\lambda_i}$ indicates the importance of the corresponding join. \Cref{eq:gnn-upd} becomes:

\begin{equation}
\begin{split}
\vec h^{(\ell)}(v) = \upd(
\vec h^{(\ell-1)}(v)
\,, 
\,
&\mu( \A^{(\ell-1)}(v,\lambda_1,o_1), m_{\lambda_1}),\dots, \\
&\mu(\A^{(\ell-1)}(v,\lambda_q,o_q), m_{\lambda_q})
)
\end{split}
\notag
\end{equation}
For each edge type $\lambda_i$, the same mask component $m_{\lambda_i}$ is applied at the aggregate vector of messages irrespective of the direction $o_i$, as both directions correspond to the same join. Also, the same mask components are used across all layers $\ell$, because message-passing corresponds to the same join operation irrespective of the GNN layer. 
We replace each aggregate with a zero replacement vector $\vec u$.

\introparagraph{(3) $\select$}
First, consider the case of a single selection predicate $\phi$. We assign a mask  $m_{\phi} \in [0,1]$ to all tuples that satisfy $\phi$. For each node $v = (R, t)$ with $\phi(t) = 1$, we set $m_{(R, t)} = m_{\phi}$.
Masking is then applied per tuple by adapting~\Cref{eq:encoder} into:
\begin{equation}
\X_v= \mu( \enc_{R} \left(\enc_{R.A_1}(t[A_1])\oplus
\cdots\oplus
\mathrm{Enc}_{R.A_n}(t[A_n]) \right), m_{(R, t)})
\notag
\end{equation}
Similarly to $\proj$, when masking $v = (R, t)$, we replace $\X_{v}$ with $\vec u = \X_{v'}$ for a replacement $v' = (R, t')$, $t' \in D(R)$ according to a random permutation of tuples in $D(R)$.

For the implementation of $\select$, we consider conditions that involve disjunctions of atomic predicates $\phi_1 \vee \dots \vee \phi_l$. Following real-valued logic techniques that map logical formulas into differentiable form~\cite{li2019consistency}, we employ the Łukasiewicz t-conorm for disjunctions. Thus, the mask values per tuple are calculated as
$m_{(R, t)} = \min\!\left\{1, \sum_{\phi_i(t, \vec c_i) = 1} m_{\phi_i} \right\}$. A similar strategy may be applied for conjunctions. 
In our experiments (\Cref{sec:experiments}), we restrict ourselves to categorical attributes with equality predicates and 
numerical attributes with range predicates defined via quantiles. 

In all cases, continuous mask values are mapped into discrete explanations using a threshold $\delta > 0$; $m'_{x} = 1$ if $m_{x} \geq \delta$ and $0$ otherwise. Components $x$ with $m'_{x} = 1$ are included in the explanations; the rest are omitted.

\section{Experimental Evaluation}
\label{sec:experiments}

\introparagraph{Metrics}
The main metric is deviation from determinacy $\devi_\Delta(E)$ (see \Cref{def:soft_determinacy}). The distance $\dist$ is absolute difference for binary classification and normalized absolute difference $\dist(x,y) = \frac{|x - y|}{|x| + |y|}$ for regression, both bounded in $[0,1]$. Each $\devi_\Delta(E)$ estimate is over 5 perturbation samples; we report both mean and standard deviation. Second, we evaluate the explanation size $k$. Recall that this is the number of projected data attributes for $\proj$, the number of join conditions for $\join$, and the number of predicates for $\select$. Finally, we report the time to generate the explanation, which corresponds to training time for mask learning.

\introparagraph{Data}
We use RelBench, the standard benchmark for RDL~\cite{relbench} with real-world databases and diverse predictive tasks. We focus on node-level tasks, i.e., regression and binary classification. 
Dataset details are provided in the online appendix. As \Cref{tab:reduce-retrain} shows, we denote each combination of database $i$ and task $j$ as D$i$T$j$.

\introparagraph{Explained Models}
We train RDL models from scratch. 
For each database, we construct a graph (\Cref{sec:prelim_rdl}) and use PyTorch Frame~\cite{hu2024pytorch_frame} for feature encoders. 
The GNN is implemented in PyG~\cite{fey2025pyg} with 32-dimensional channels. We perform hyperparameter tuning over the learning rate, batch size, and number of GNN layers, and use the train/val/test split of the benchmark. Our models reach performance similar to the one previously reported~\cite{relbench} for GNN approaches.  
For each explanation task, we sample 100 instances for mask training and 100 instances for evaluation. 
\rebuttal{R1D1}{For classification tasks, we report a class-balanced $dev_\Delta$ that averages over an equal number of instances from each class, avoiding majority-class dominance.}
For temporal tasks we perform temporal sampling \cite{fey2024position} to ensure time-consistency.

\introparagraph{Methods}  
(1) Mask-based:  
Our method as introduced in \Cref{sec:model_specific}, with early stopping \rebuttal{R1D4}{and thresholding with $\delta=0.1$. Note that, on average, \emph{only} 1.97\% of mask values for $\proj$ and 2.17\% for $\join$ fall in the range $[0.05, 0.2]$ and are sensitive to small $\delta$ variations}.  
(1a)~\colmask: Masks for $\proj$. (1b)~\fkpkmask: Masks for $\join$. (1c)~\filtermask: Masks for $\select$. 

Since no prior work generates SQL explanations for RDL, we implement a set of baselines.  
(2)~Ranking-based for $\proj$: 
We order data attributes according to their importance, and select the top-$k$ for each target explanation size $k$. A known limitation of such methods is that they ignore dependencies between attributes \cite{molnar2020interpretable}.  
(2a)~\greedysubset: The importance of a data attribute is computed as $dev_\Delta$ when the attribute is taken as an explanation by itself. The lower the $dev_\Delta$ the more important the attribute. (2b)~\pfi: We adapt Permutation Feature Importance (PFI) \cite{molnar2020interpretable} by measuring the change in $dev_\Delta$ when each attribute is removed. Attributes are ranked by impact, where higher deviation implies higher importance.  

(3)~Greedy approach for $\proj$ and $\join$:  
(3a)~\greedysubsetiterative: Starting from the empty set, we iteratively add the attribute that yields the largest reduction in $dev_\Delta$. (3b)~\greedyexpansion: At each step, we add a join that maximally reduces $dev_\Delta$ while ensuring connectivity to the prediction entity. This is more computationally feasible than \greedysubsetiterative\ due to fewer join candidates at each iteration.

(4)~\randomsubset\ for $\proj$: We randomly sample the data attributes and report the average over 5 samples.

\introparagraph{Experiment details} 
For text attribute embeddings we use a BERT-based model\footnote{\url{https://huggingface.co/sentence-transformers/distilbert-base-nli-mean-tokens}}.
Experiments are done on a VM with an RTX 6000 Ada GPU, 14 vCPUs, and 188 GB of RAM. Our code is public\footnote{\url{https://github.com/agapiR/rdl-explain}}.

\introparagraph{Evaluation plan}
We evaluate mask learning separately for $\proj$ and $\join$. For fair comparison, we use the same explanation size for all methods, denoted as $k^*$, which we determine by mask thresholding as discussed in~\Cref{sec:model_specific}.
We also assess the robustness of $dev_\Delta$ under different database perturbation strategies $\Delta$. Finally, we conduct a case study to show end-user usefulness.

\subsection{$\proj$ Evaluation}

\begin{figure*}[ht]
    \centering
    \begin{subfigure}[b]{0.4\textwidth}
        \centering
        \includegraphics[width=\textwidth]{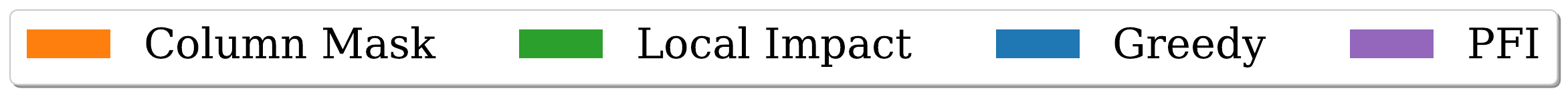}
    \end{subfigure}
    \vspace{-0.8mm}
    
    \begin{subfigure}[b]{\textwidth}
        \centering
        \includegraphics[width=\textwidth]{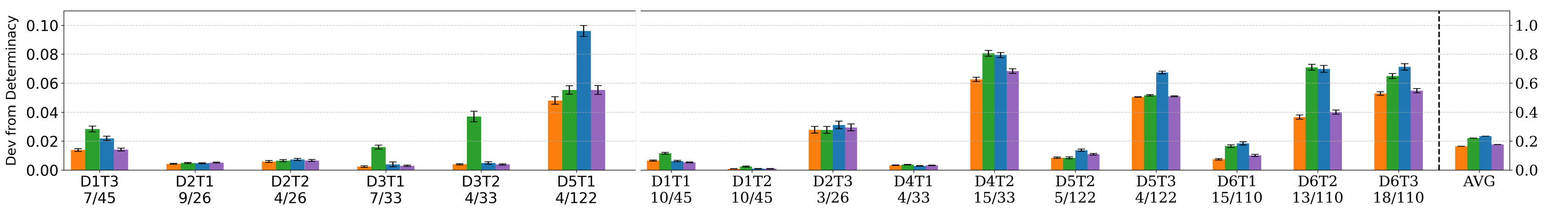}
        \caption{Deviation from determinacy $\devi_\Delta$ for each database-task and database perturbation $\Delta^{\mathsf{ind}}_{E}$.}
        \label{fig:proj_dev_independent}
    \end{subfigure}
        
    \begin{subfigure}[b]{\textwidth}
        \centering
        \includegraphics[width=\textwidth]{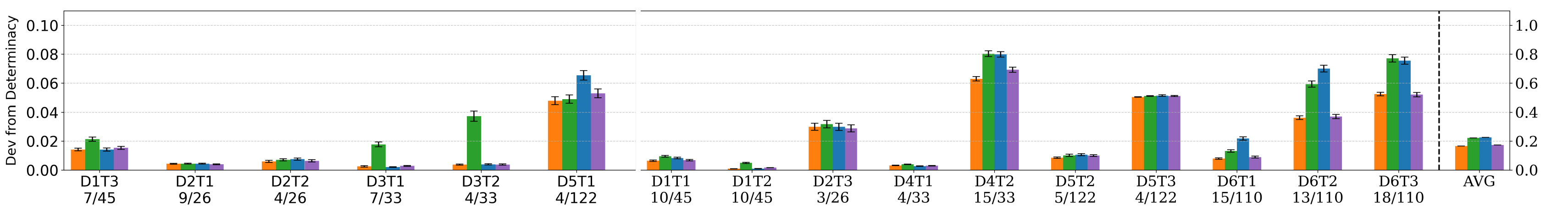}
        \caption{Deviation from determinacy $\devi_\Delta$ for each database-task and database perturbation $\Delta^{\mathsf{joint}}_{E}$.}
        \label{fig:proj_dev_joint}
    \end{subfigure}

    \begin{subfigure}[b]{\textwidth}
        \centering
        \includegraphics[width=\textwidth]{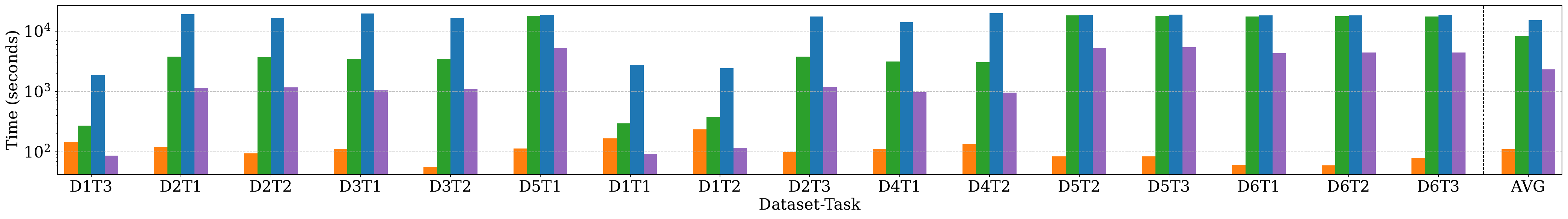}
        \caption{Time to obtain explanation for each database-task and $\Delta^{\mathsf{ind}}_{E}$. Results for $\Delta^{\mathsf{joint}}_{E}$ are similar.}
        \label{fig:proj_time}
    \end{subfigure}
    
    \caption{Evaluation for $\proj$. For visualization purposes, tasks are separated into ``easy'' (left) and ``hard'' (right) ones, with separate $\devi_\Delta$ scales. The average (AVG) across all tasks is also shown. Below the dataset and task name, we indicate the explanation size $k^*$ and the total number of data attributes.}
    \label{fig:proj-all}
\end{figure*}

\Cref{fig:proj-all} reports the results for $\proj$ across all datasets and tasks, including averages (AVG).\Cref{fig:proj_dev_independent} and \Cref{fig:proj_dev_joint} show $\devi_\Delta$ for two different perturbation strategies. The main takeaway is that \colmask\ produces higher quality explanations (in terms of $\devi_\Delta$) than the baselines on most tasks, and on average, while outperforming them on execution time (\Cref{fig:proj_time}) by 1-2 orders of magnitude for large databases. This is expected since \colmask's running time is dependent on the size of the model and the size of the explanation task, which is usually small. 
Among the baselines, \pfi\ gives the best results both on quality and on time.
However, all baselines suffer from their costly dependence on the schema size, i.e., the number of data attributes that they consider.
On database D$1$ \pfi\ is faster, but this is due to the fact that the database is small and has a concise schema. Comparing the two perturbation strategies, they appear to yield very similar results, indicating that $\devi_\Delta$ is robust to the choice of permutation strategy.

\begin{figure*}[t]
    \centering

    \includegraphics[width=0.5\textwidth]{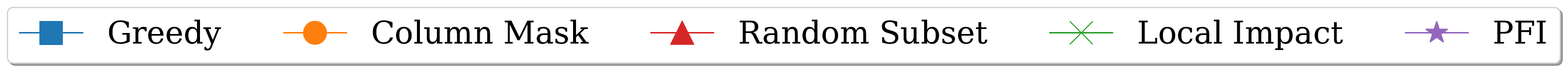}
    \vspace{-0.5mm}

    \begin{subfigure}[b]{0.22\textwidth}
        \centering
        \includegraphics[width=\linewidth]{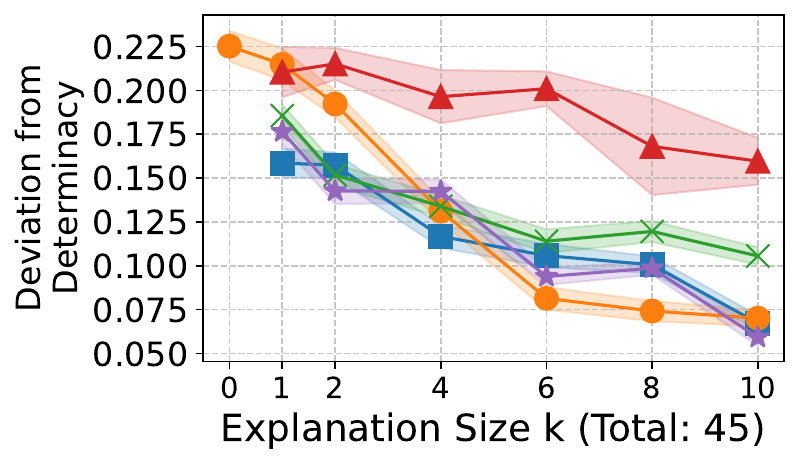}
        \caption{D1T1}
    \end{subfigure}
    \begin{subfigure}[b]{0.22\textwidth}
        \centering
        \includegraphics[width=\linewidth]{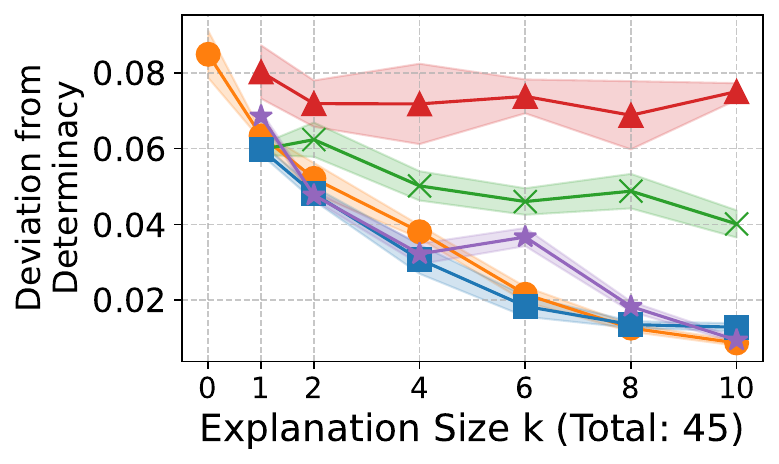}
        \caption{D1T2}
    \end{subfigure}
    \begin{subfigure}[b]{0.22\textwidth}
        \centering
        \includegraphics[width=\linewidth]{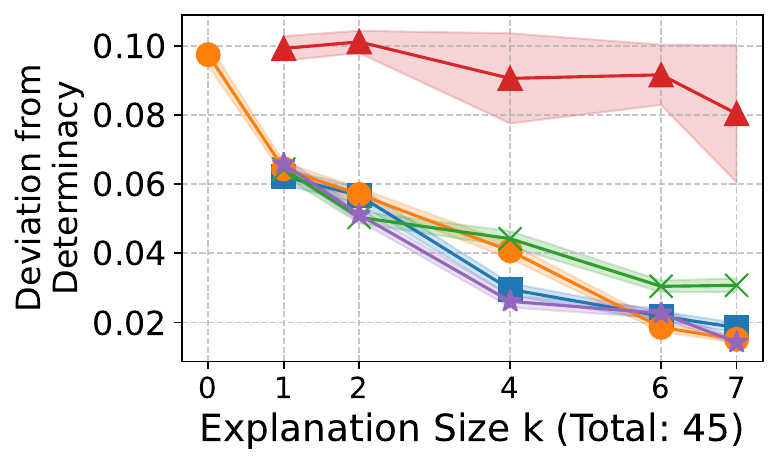}
        \caption{D1T3}
    \end{subfigure}
    \begin{subfigure}[b]{0.22\textwidth}
        \centering
        \includegraphics[width=\linewidth]{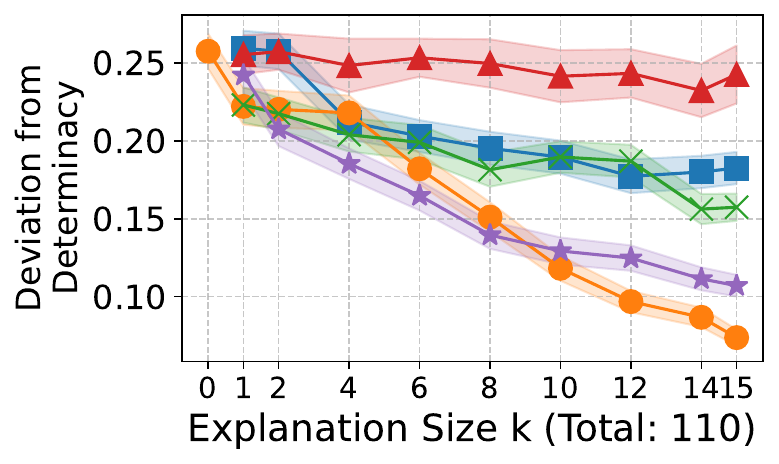}
        \caption{D6T1}
    \end{subfigure}

    \begin{subfigure}[b]{0.22\textwidth}
        \centering
        \includegraphics[width=\linewidth]{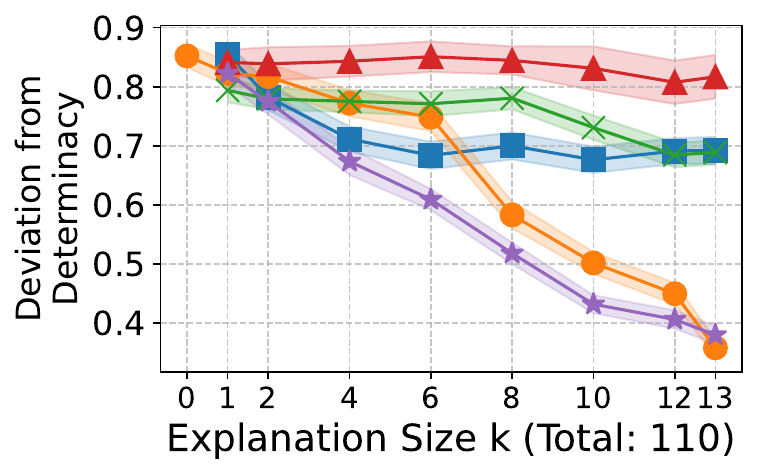}
        \caption{D6T2}
    \end{subfigure}
    \begin{subfigure}[b]{0.22\textwidth}
        \centering
        \includegraphics[width=\linewidth]{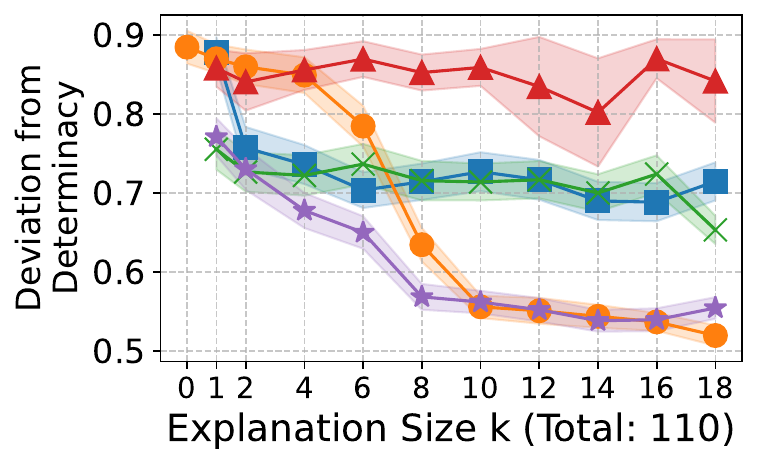}
        \caption{D6T3}
    \end{subfigure}
    \begin{subfigure}[b]{0.22\textwidth}
        \centering
        \includegraphics[width=\linewidth]{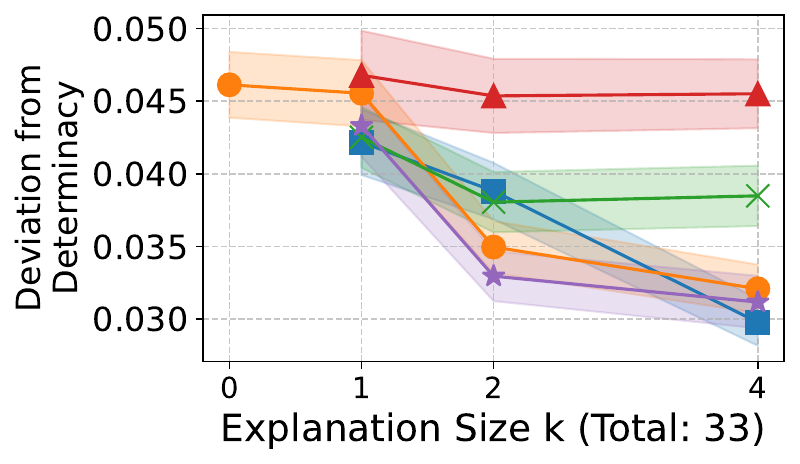}
        \caption{D4T1}
    \end{subfigure}
    \begin{subfigure}[b]{0.22\textwidth}
        \centering
        \includegraphics[width=\linewidth]{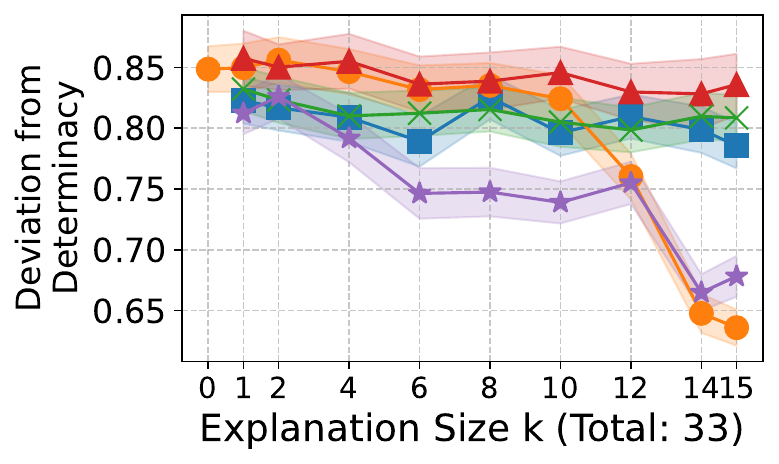}
        \caption{D4T2}
    \end{subfigure}

    \caption{Soft determinacy versus conciseness for $\proj$ explanations, for $8$ tasks and database perturbation $\Delta^{\mathsf{ind}}_{E}$.
    }
    \label{fig:proj-curves}
\end{figure*}

While \Cref{fig:proj-all} assumes the same explanation size $k^*$, \Cref{fig:proj-curves} shows 
$dev_\Delta$ for various explanation sizes $k$, illustrating the tradeoff between the two.
As the size $k$ increases, we generally observe $dev_\Delta$ decreasing, i.e., improving explanation quality.
This is not true for \randomsubset, which shows that the choice of attributes in the explanation is crucial.
For smaller $k$ than $k^*$, the baselines are often better than \colmask, 
but \colmask\ usually outperforms them for the optimal $k^*$ (i.e., the end of the axis).

\subsection{$\join$ Evaluation}
\label{sec:fkjoin_eval}
Our method \fkpkmask\ achieves an average (across all datasets and tasks) $\devi_\Delta$ of $0.1366 \pm 0.0020$ for $\Delta^{\mathsf{uniform}}_{E}$ and $0.1359 \pm 0.0022$ for $\Delta^{\mathsf{freq}}_{E}$. The baseline \greedyexpansion\ yields $0.1400 \pm 0.0022$ and 
$0.1415 \pm 0.0023$ for the respective distributions. In terms of $\devi_\Delta$, the gap between \fkpkmask\ and \greedyexpansion\ is smaller than in previous experiments, but \fkpkmask\ still edges out on average, winning 53\% of the time (15 out of 28 comparisons). As before, the mask-based approach \fkpkmask\ is significantly faster than the greedy baseline \greedyexpansion, requiring only 101.46 seconds on average compared to 2114.02 seconds under $\Delta^{\mathsf{uniform}}_{E}$ and 2844.50 seconds under $\Delta^{\mathsf{freq}}_{E}$, offering a speedup of 1-2 orders of magnitude consistent across the vast majority of datasets-tasks. The differences between the two perturbation strategies are again minor, verifying the robustness of our metric.

\rebuttal{R1W2 and R1D9c}{
\subsection{Case Study: Diagnosing Model Behavior}
\label{sec:exp_case_study}

We demonstrate how our framework's expressivity enables users to diagnose model behavior through a series of controlled scenarios on the rel-trial dataset \cite{clinical2019improving}, a database of clinical trial reports. It contains 15 tables with a total of 5,852,157 rows and 140 attributes, out of which 110 are data attributes and 30 are key or foreign-key attributes. We focus on the binary classification study-outcome (D6T1), that predicts whether a trial achieves its primary outcome. To show how different $\exlang$s reveal distinct issues, we also construct 3 synthetic variants of the original task, summarized in \Cref{tab:case-study-summary}.

\introparagraph{Original Task}
Our method identifies the deciding factors for whether a study will achieve its outcome. It highlights ${\sim}10\%$ of columns and $10$ foreign-key pairs using $\proj$ and $\join$ respectively. 
Utilized facilities (``\sql{studies JOIN  facilities\_studies JOIN facilities}'') and study designs (``\sql{studies JOIN  designs}'') are revealed as key predictors. 
The relation \sql{designs} originally contains $13$ data attributes, of which $3$ are important: \sql{allocation}, \sql{intervention\_model}, \sql{primary\_purpose}. 
The relation \sql{facilities} contains $3$ additional important data attributes: \sql{name}, \sql{city}, and \sql{country}, indicating that names and locations of utilized facilities highly influence the study success or failure.
Notably, \emph{no} attributes from the \sql{outcomes} relation (e.g., outcome type or description) are included, indicating they do not help predict outcome achievement. 
We further examine the important selection predicates discovered by our method using $\select$. Our method detects the following predictive predicates: ``\sql{designs.allocation = 0}'' selects single-arm trials, ``\sql{designs.intervention\_model = 0}'' selects single-group interventions and ``\sql{studies.phase $\in \{0,2,4\}$}'' selects trials based on the phase. 
}

\rebuttal{}{
\introparagraph{Scenario 1: Detecting Column-Level Data Leakage}
We simulate a realistic mistake: a user inadvertently includes a \sql{prelim\_evaluation} table where \sql{studies} receive an evaluation from multiple reviewers (many-to-many join) and the numerical column \sql{rating} ($\in \{1,2,3,4,5\}$) as part of the evaluation record is perfectly correlated with \sql{outcome} (high grades $(\in \{4,5\})$ $\rightarrow$ positive outcomes, low grades $(\in \{1,2\})$ $\rightarrow$ negative outcomes). 
The trained model achieves suspiciously high performance (ROC-AUC $\simeq 100\%$). Using $\proj$, our method immediately flags the \sql{rating} column as highly important ($\text{mask}_\text{\sql{rating}} \simeq 1$), while mask values for all other columns are $\simeq 0$. 
This provides a clear signal to investigate potential leakage in this specific column.
}

\rebuttal{}{
\introparagraph{Scenario 2: Pinpointing Tuple-Level Leakage}
We now make the leakage more subtle: each evaluation record refers to a certain evaluation aspect (e.g., ``financial'', ``ethics'', ``methodology'' etc.) indicated by a column \sql{eval\_category} that admits $10$ categorical values (``A'', $\dots$, ``J''). In this case, only ratings on the first three categories (``A'', ``B'', ``C'') are correlated with outcomes (as before) while the rest are uncorrelated. Model performance is again high. $\proj$ identifies \sql{rating} as important along with \sql{eval\_category}. Subsequently, to understand \emph{which} evaluations cause leakage, we use $\select$ on the \sql{eval\_category} column. The method precisely identifies the predicate \sql{eval\_category IN (``A'', ``B'', ``C'')} with high mask value, while other evaluation records have low importance. This fine-grained diagnosis enables targeted data cleaning.
}

\rebuttal{}{
\introparagraph{Scenario 3: Understanding Structural Importance}
We make the outcome depend purely on structure: \sql{outcome = 1} if the study has multiple sponsors, else \sql{0}. The model learns this pattern well. $\join$ explanations reveal the \sql{sponsors\_studies} join as critical. However, $\proj$ explanations show that \emph{no specific sponsor attributes} are important, the predictive signal comes entirely from the join structure (i.e., counting sponsors). This illustrates that expressivity across multiple languages is essential: neither $\proj$ nor $\join$ alone would provide the complete picture.
}

\begin{table}[t]
\centering
\footnotesize
\caption{Summary of case study scenarios showing how different explanation languages reveal distinct patterns.}
\label{tab:case-study-summary}
\begin{tabular}{@{}lccc@{}}
\toprule
\textbf{Scenario}  & $\exlang$ & \textbf{Size} & \textbf{Key Artifacts} \\
\midrule
Original  & \textsc{ALL} & 20 & facilities, designs, etc. \\
Column Leakage  & $\proj$ & 1 & \sql{SELECT rating FROM prelim\_evaluation} \\
Tuple Leakage  & $\select$ & 3 & \sql{eval\_category IN (``A'', ``B'', ``C'')} \\
Structural  & $\join$ & 1 & \sql{studies JOIN sponsors\_studies} \\
            & $\proj$ & 0 & No sponsor data attributes \\
\bottomrule
\end{tabular}
\end{table}

\subsection{Retraining Results}

As a sanity check, we retrain the models using only the subset of the data retained by our $\proj$ and $\join$ explanations. As \Cref{tab:reduce-retrain} shows, performance is comparable for most tasks while the database size is substantially smaller. This confirms that the selected data retain the core predictive signal.

\section{Related Work}
\label{sec:related_work}

\introparagraph{Learning over a relational database} Statistical relational learning \cite{getoor2007srl, raedt2016statistical} is one of the first approaches to directly leverage the relational structure, factorized learning \cite{schleich_olteanu_2016, kumar2016join, relational_borg} avoids costly joins, automatic feature engineering \cite{kanter2015DFS_deepfeaturesynthesis, arda, zhang2023gfs, stanoev2024automating} synthesizes new features, while other approaches train arbitrary (often tree-based \cite{xgboost}) predictive models. 
The currently emerging approach is Relational Deep Learning (RDL) \cite{cvitkovic2020supervised, fey2024position, 4dbinfer_wang2024}. It models the relational database as a heterogeneous graph, and employs tabular feature encoders \cite{hu2024pytorch_frame} trained jointly with Graph Neural Networks (GNNs) to solve a wide range of predictive tasks. 
The current research frontier further builds specialized architectures for RDL that fall under the same GNN paradigm \cite{hilprecht2023spare, chen2025relgnn, yuan2025contextgnn} and adapts Graph Transformers to the relational setting \cite{rampavsek2022recipe, kong2023goat, dwivedi2025relational_graph_transformer}. Of course, the RDL approach inherits the opacity of ``black-box'' deep learning, creating the need for explainability. Since RDL models, irrespective of their implementation, make predictions at the database level, we consider explanations that refer to the database rather than the 
specific modeling components, such as the heterogeneous graph.

\introparagraph{ML explanations} 
To address the explainability in RDL, we review the broader XAI literature. Explanations are categorized as either \e{local} (i.e., focusing on individual predictions) or \e{global} (i.e., explaining a model's overall behavior) \cite{guidotti2019survey, samek2019explainable, molnar2020interpretable, bassan24_local_global}. For example, local feature attribution methods \cite{ribeiro2016should_lime, lundberg2017unified, sundararajan2017axiomatic_IG}, explain individual predictions by assigning importance scores to each feature associated with a specific instance, while global feature importance methods like PFI \cite{breiman2001_pfi} and its variants \cite{fisher2019_pfi, debeer2020cpi, molnar2024conditional_subgroups}, detect important features across all predictions by measuring the each feature's impact on model performance.
A more formal distinction further classifies these methods into \e{abductive} and \e{contrastive} \cite{marques2022delivering, barcelo2025explaining}. Abductive explanations identify minimal subsets of the input \e{sufficient} to preserve the model's prediction, whereas contrastive explanations specify the minimal changes required to alter it. In our context, explanation views are global, as they refer to the database as a whole, and abductive, as they identify the database components sufficient to determine the model's behavior. 
\rebuttal{R1D9d, R3C2 and R3D2}{
Crucially, traditional XAI assumes a feature vector associated with each prediction instance. In RDL, there is no single ``prediction tuple'' containing all relevant features. Instead, predictions use the relational graph where tuples across multiple tables are connected, and features are derived through message-passing and aggregation. This makes these methods conceptually misaligned with RDL, as they ignore that features are computed by combining multiple connected tuples and that the relational structure itself carries predictive information}.

\introparagraph{Subgraph-based GNN explanations}
\emph{Instance-level} (i.e., local) explanations specific to Graph Neural Networks (GNNs) aim to answer questions of the type ``Why does the GNN produce a particular output \emph{for a given instance} (i.e., node in the graph)?''. 
Most of these explanations are at the level of individual graph elements, i.e., the explanation is a subgraph of the input graph. 
Some instance-level explainers focus on neighboring nodes and their individual features \cite{pope2019explainability_methods, vu2020pgm-explainer, funke2021hard_masking, li2023heterogeneous}, others on important edges \cite{ying2019gnnexplainer, lucic2022cf}, while others on connected subgraphs \cite{yuan2021subgraphx, lin2021generative, shan2021reinforcement, li2023dag, mika2023hgexplainer}. 
Few are tailored to hetero-GNNs \cite{li2023heterogeneous, mika2023hgexplainer} by specializing subgraph explanations to meta-paths.
Similarly, \emph{global} GNN explainers, which provide insights for a GNN across all instances, use subgraphs as explanations. 
More specifically, some recover a collection of subgraphs that collectively explain many instances \cite{xgnn, data_aware, wanggnninterpreter, huang2023global, chen2024view} aiming at explanation generality, while others extract instance-level explanation subgraphs which are then combined into global insights \cite{glocal, azzolin2023global, chen2024view}. In contrast to our approach, most global explainers implicitly assume the predictive tasks are motif-based \cite{azzolin2025beyond_topological} and they are often evaluated against a ground truth. We refer the reader to a recent survey for more details on these techniques \cite{yuan2022taxonomy}. We depart from all these subgraph-based approaches in the following ways: 
\circled{1}~Our explanations are expressed in terms of the relational database, which is the input object, instead of the graph that is part of modeling. 
\circled{2}~The complexity of our explanations is measured on SQL views, which can succinctly refer to billions of data points. 
This implies that subgraph-based explanations can be expressed as SQL views, often much more succinctly than in graph-based representations.
\circled{3}~Our SQL explanations can express database components that go beyond simple subgraphs. 
For example, a single-column projection does not correspond to a connected subgraph. It is also possible for explanations to refer to derived information, such as aggregates which do not explicitly exist in the graph at all.

\introparagraph{Other GNN explanations}
Beyond subgraphs, there are other types of explanations that are fundamentally different. First, some approaches, if adapted to RDL, would not correspond to any database elements \cite{vu2020pgm-explainer, zhang2021relex, huang2022graphlime, baldassarre2019explainability, schnake2021higher, schwarzenberg2019layerwise, schlichtkrull2021graphmask}, e.g., because they refer to GNN layers or construct surrogate models. Second, distillation \cite{pluska24_distill} can map the GNN into a logical classifier. This is similar in spirit to our explanations; however, their base predicates refer to inner features of the model and can quickly become too complex. Finally, self-explainable GNNs \cite{dai2021selfexplain} claim to provide simpler GNN models, thus with better interpretability properties. Recently this approach has been adapted to RDL, but it is limited to meta-path selection \cite{ferrini2025_self_explainable}. Our explanations are given post-hoc on an already trained model, instead of training a simpler model from scratch. 

{\begin{table}[t]
\centering
\footnotesize  %
\caption{Databases, tasks, model performance, and performance after training with only the data in our explanations.
}
\label{tab:reduce-retrain}
\vspace{-4mm}
\begin{tabular}{lllcccc}
\toprule
\textbf{Database} & \textbf{Task} & \textbf{D$i$T$j$} & \textbf{Perf.} & \textbf{\makecell{Masked\\Perf.}} & \textbf{Diff} & \textbf{Size} $\boldsymbol{\downarrow}$ \\
\midrule
\multicolumn{7}{c}{\textbf{Binary Classification (ROC-AUC)}} \\
\midrule
rel-f1 & driver-dnf & D1T1 & 75.91 & 73.27 & -2.64 & 51.19\% \\
rel-f1 & driver-top3 & D1T2 & 76.50 & 67.29 & -9.21 & 55.49\% \\
rel-avito & user-clicks & D2T1 & 65.03 & 67.28 & 2.26 & 1.93\% \\
rel-avito & user-visits & D2T2 & 64.45 & 63.64 & -0.81 & 8.34\% \\
rel-stack & user-engage & D3T1 & 89.92 & 89.60 & -0.32 & 64.56\% \\
rel-stack & user-badge & D3T2 & 88.33 & 88.11 & -0.23 & 84.05\% \\
rel-hm & user-churn & D4T1 & 68.79 & 68.82 & 0.03 & 50.84\% \\
rel-trial & study-outcome & D6T1 & 71.33 & 69.40 & -1.92 & 79.46\% \\
rel-event & user-repeat & D5T1 & 75.45 & 78.60 & 3.16 & 94.81\% \\
rel-event & user-ignore & D5T2 & 79.34 & 75.22 & -4.12 & 73.21\% \\
\midrule
\multicolumn{7}{c}{\textbf{Regression (MAE)}} \\
\midrule
rel-f1 & driver-position & D1T3 & 3.2444 & 3.3487 & 0.1043 & 51.22\% \\
rel-avito & ad-ctr & D2T3 & 0.0451 & 0.0457 & 0.0006 & 11.86\% \\
rel-hm & item-sales & D4T2 & 0.0574 & 0.0566 & -0.0008 & 31.35\% \\
rel-trial & study-adverse & D6T2 & 46.7445 & 45.4555 & -1.2890 & 85.23\% \\
rel-trial & site-success & D6T3 & 0.3245 & 0.3747 & 0.0502 & 76.02\% \\
rel-event & user-attendance & D5T3 & 0.2449 & 0.2381 & -0.0067 & 94.51\% \\
\bottomrule
\end{tabular}
\end{table}
}

\introparagraph{Success metrics}
Most prior work focuses on sufficiency of explanations, i.e., finding a small subgraph that (by itself) retains the GNN prediction, quantified by a metric called \emph{fidelity}\footnote{Fidelity is sometimes referred to as fidelity minus \cite{amara2022graphframex} or sufficiency \cite{longa2025explaining}} \cite{yuan2022taxonomy}. 
Variants of this metric either focus on retaining prediction confidence \cite{li2023heterogeneous} or robustness under random,  \cite{zheng2024towards_robust}, infinitesimally small \cite{agarwal2022probing}, and even adversarial \cite{fang2023evaluating_robustness} perturbations. 
In addition to sufficiency, some works assess explanation necessity \cite{amara2022graphframex, longa2025explaining}, i.e., whether the explanation subgraph is essential to maintain the prediction. 
Closest to soft determinacy is a fidelity variant called degree of sufficiency \cite{azzolin2025reconsidering, azzolin2025beyond_topological} in that it has a probabilistic nature. 
Compared to all this prior work, our evaluation has the following key differences: 
\circled{\normalsize{1}}~ Since the framework is general, soft determinacy is parameterized by the explanation language $\exlang$. In contrast to subgraph modifications, the perturbations needed for soft determinacy are in the database space and must respect $\exlang$ to be valid. 
\circled{\normalsize{2}}~ We minimize explanation complexity with respect to query instead of data size.
\circled{\normalsize{3}}~ We also focus on the time to obtain an explanation, which has received limited attention in prior work, with few exceptions \cite{yuan2021subgraphx, lin2021generative, chen2024view}.

\introparagraph{Explanations in databases}
In the context of database research, explanations target various phenomena including why specific query results are present or not (why-not explanations) \cite{whynot}, outliers \cite{miao2019cape_outliers, wu2013scorpion_outliers}, trends \cite{roysuciu2014} or bias \cite{Salimi23} in query outputs, and performance anomalies \cite{kalmegh2016iQCAR, zhang2017exstream}. 
\rebuttal{R2D3}{
Diverse approaches have been employed: Provenance-based methods \cite{Glavic21} attribute query results back to input tuples, with techniques ranging from deriving provenance expressions \cite{buneman2001whyandwhere, greentannen2017semiring, Amsterdamer11, Glavic2013, Senellart18} to quantifying tuple contributions via Shapley values \cite{Livshits20, kimelfeld21query_games, Deutch22} or responsibility measures \cite{Meliou10, Meliou11, Freire15}. Intervention-based approaches perform targeted data perturbations (e.g., tuple deletion or addition) to assess causal impact on particular outcomes \cite{wu2013scorpion_outliers, roysuciu2014, Roy15a}. 
Summarization methods place a focus on conciseness, e.g., employing predicates \cite{wu2013scorpion_outliers, roysuciu2014, Gebaly14, Roy15a, gebaly2018explanation_tables, Lee20, Abuzaid} or taxonomies \cite{Wang15a, Glavic15} to succinctly describe significant parts of the database. 
Example-based techniques focus on minimal examples to enhance interpretability, e.g., highlighting errors via counterexamples \cite{Miao19}. These approaches are not mutually exclusive and can be combined, e.g., using summarization to refine complex provenance expressions \cite{Cate15, Lee20}. A complete taxonomy can be found in \cite{glavic2021trends}.}
These efforts parallel XAI in aiming to explain algorithmic outputs via input data and/or computational logic. 
Differently, in our framework, explanations are themselves simple queries, explaining  complex queries (deep models).

\section{Conclusions and Future Work}
\label{sec:conclusion}

We presented a model-agnostic framework for abductive global explanations of ML models over relational databases, 
and studied its instantiation on hetero-GNNs, for which we proposed an efficient mask-learning approach. Our extensive evaluation on the RelBench benchmark 
demonstrated that our method discovers high-quality explanations with low runtime across diverse tasks.

Several directions for future work emerge from our framework. A first class of extensions concerns broadening the applicability and expressivity. 
\rebuttal{R1W3 and R1D9b}{
One natural direction is to support additional explanation languages using aggregates, grouping, nesting, or more complex selection predicates. To achieve this, several interesting challenges arise, including defining appropriate database perturbations that respect the semantics of these constructs, systematically identifying good candidate predicates, and investigating efficient approaches for explanation discovery, since it is unclear whether and how mask-learning can be applied.}
Another direction is to develop efficient instantiations for different model classes, e.g., attention-based models~\cite{DBLP:journals/corr/abs-2505-05568,hollmann2025tabpfn,DBLP:journals/corr/abs-2506-16654}, and different predictive tasks, e.g., recommendations. Beyond these, a fundamental question is to identify explanation languages that naturally align with a given use case; can we predict which explanation constructs will yield the most useful results for a particular combination of model architecture and database schema?

\clearpage

\bibliographystyle{ACM-Reference-Format}
\bibliography{bibliography}

\newpage
\appendix
\section{Experimental Details}

\paragraph{Benchmark Details}
\Cref{tab:benchmark_summary_detailed} provides a summary for the Relbench Benchmark including dataset statistics and our best models' performance in comparison to SOTA performance. 

\paragraph{Hyperparameters for Mask Learning} 
Masks were optimized using Adam optimizer \cite{KingmaB15} with default settings, except for the learning rate, which was fixed to $0.05$ across all experiments. We employed early stopping based on convergence of a smoothed version of the mask learning objective: optimization was terminated when the absolute difference between the mean loss over two consecutive sliding windows of five epochs fell below $10^{-6}$. For each dataset and task, we selected the regularization coefficient $\lambda$ to balance the initial task loss and the mask regularization term at the start of mask learning. The chosen $\lambda$ values are reported in \Cref{tab:hyperparams}, separately for $\proj$ and $\join$. We additionally report the number of epochs required to reach convergence for each configuration. After mask learning, we employ mask discretization via thresholding with threshold $\delta=0.1$.

\begin{table*}[t]
\centering
\caption{Datasets, tasks, and performance comparison with SOTA \cite{relbench, lachi2025boosting}. Database-level statistics are shared across tasks. We report our best performance achieved in each task. The models we explain in the main paper have similar performance.}
\label{tab:benchmark_summary_detailed}
\footnotesize
\vspace{-2mm}
\begin{tabular}{llcccccll}
\toprule
\textbf{Database} & \textbf{D$i$T$j$} & \textbf{Task} & \textbf{\#Tuples} & \textbf{\#Attr.} & \textbf{\#Data Attr.} & \textbf{Size (MB)} & \textbf{Perf. (Ours)} & \textbf{Perf. (SOTA)} \\
\midrule
\multicolumn{9}{c}{\textbf{Binary Classification (ROC-AUC)}} \\
\midrule
\multirow{2}{*}{rel-f1}
  & D1T1 & driver-dnf   & \multirow{2}{*}{97,606} & \multirow{2}{*}{67} & \multirow{2}{*}{45} & \multirow{2}{*}{1.18} & 0.7777 & 0.7355 \\
  & D1T2 & driver-top3  &                         &                     &                     &                        & 0.7677 & 0.8473 \\
\midrule
\multirow{2}{*}{rel-avito}
  & D2T1 & user-clicks  & \multirow{2}{*}{24,653,915} & \multirow{2}{*}{42} & \multirow{2}{*}{26} & \multirow{2}{*}{443.83} & 0.6622 & 0.6590 \\
  & D2T2 & user-visits  &                           &                      &                      &                           & 0.6445 & 0.6620 \\
\midrule
\multirow{2}{*}{rel-stack}
  & D3T1 & user-engagement & \multirow{2}{*}{5,399,818} & \multirow{2}{*}{52} & \multirow{2}{*}{33} & \multirow{2}{*}{1004.46} & 0.8992 & 0.9059 \\
  & D3T2 & user-badge      &                           &                      &                      &                            & 0.8833 & 0.8886 \\
\midrule
\multirow{1}{*}{rel-hm}
  & D4T1 & user-churn & 16,931,173 & 37 & 33 & 179.99 & 0.6879 & 0.6998 \\
\midrule
\multirow{2}{*}{rel-event}
  & D5T1 & user-repeat  & \multirow{2}{*}{59,141,053} & \multirow{2}{*}{246} & \multirow{2}{*}{122} & \multirow{2}{*}{440.74} & 0.7876 & 0.7777 \\
  & D5T2 & user-ignore  &                             &                       &                       &                          & 0.7934 & 0.8398 \\
\midrule
\multirow{1}{*}{rel-trial}
  & D6T1 & study-outcome & 5,852,157 & 140 & 110 & 649.79 & 0.7133 & 0.7009 \\
\midrule
\multicolumn{9}{c}{\textbf{Regression (MAE)}} \\
\midrule
\multirow{1}{*}{rel-f1}
  & D1T3 & driver-position & 97,606 & 67 & 45 & 1.18 & 3.2444 & 4.022 \\
\midrule
\multirow{1}{*}{rel-avito}
  & D2T3 & ad-ctr & 24,653,915 & 42 & 26 & 443.83 & 0.0401 & 0.0410 \\
\midrule
\multirow{1}{*}{rel-hm}
  & D4T2 & item-sales & 16,931,173 & 37 & 33 & 179.99 & 0.0561 & 0.0560 \\
\midrule
\multirow{1}{*}{rel-event}
  & D5T3 & user-attendance & 59,141,053 & 246 & 122 & 440.74 & 0.2404 & 0.2580 \\
\midrule
\multirow{2}{*}{rel-trial}
  & D6T2 & study-adverse & \multirow{2}{*}{5,852,157} & \multirow{2}{*}{140} & \multirow{2}{*}{110} & \multirow{2}{*}{649.79} & 46.2990 & 44.0110 \\
  & D6T3 & site-success  &                             &                       &                       &                           & 0.3245 & 0.4000 \\
\bottomrule
\end{tabular}
\end{table*}

\begin{table*}
\centering
\caption{Hyperparameters for mask learning, per task. \#epochs refers to the number of epochs before early stopping, with a maximum of $250$ epochs used across all experiments.}
\label{tab:hyperparams}
\footnotesize 
\vspace{-4mm}
\begin{tabular}{lllcccc}
\toprule
\textbf{Database} & \textbf{Task} & \textbf{$\lambda$} & \textbf{\#epochs} \\
\midrule
\multicolumn{4}{c}{\textbf{$\proj$}} \\
\midrule
rel-f1 & driver-dnf & 0.01 & 250 \\
rel-f1 & driver-top3 & 0.01 & 250 \\
rel-f1 & driver-position & 0.1 & 250 \\
rel-avito & user-clicks & 0.0001 & 191 \\
rel-avito & user-visits & 0.0001 & 250 \\
rel-avito & ad-ctr & 0.001 & 250 \\
rel-stack & user-engagement & 0.0001 & 250 \\
rel-stack & user-badge & 0.001 & 250 \\
rel-hm & user-churn & 0.001 & 250 \\
rel-hm & item-sales & 0.001 & 250 \\
rel-event & user-repeat & 0.001 & 250 \\
rel-event & user-ignore & 0.01 & 250 \\
rel-event & user-attendance & 0.001 & 250 \\
rel-trial & study-outcome & 0.01 & 250 \\
rel-trial & study-adverse & 1 & 250 \\
rel-trial & site-success & 0.01 & 250 \\
\midrule
\multicolumn{4}{c}{\textbf{$\join$}} \\
\midrule
rel-f1 & driver-dnf & 0.001 & 250 \\
rel-f1 & driver-top3 & 0.001 & 250 \\
rel-f1 & driver-position & 0.01 & 250 \\
rel-avito & user-clicks & 0.0001 & 241 \\
rel-avito & user-visits & 0.001 & 161 \\
rel-avito & ad-ctr & 0.01 & 250 \\
rel-stack & user-engagement & 0.01 & 250 \\
rel-stack & user-badge & 0.01 & 250 \\
rel-event & user-repeat & 0.001 & 161 \\
rel-event & user-ignore & 0.01 & 250 \\
rel-event & user-attendance & 0.01 & 211 \\
rel-trial & study-outcome & 0.001 & 250 \\
rel-trial & study-adverse & 0.1 & 250 \\
rel-trial & site-success & 0.01 & 250 \\
\bottomrule
\end{tabular}
\end{table*}

\section{Composite Languages}

\begin{figure}[ht]
\centering
\begin{subfigure}{0.9\linewidth}
\begin{lstlisting}
SELECT  id, nct_id, 
        allocation, intervention_model, primary_purpose
FROM    designs d
WHERE   d.allocation=0  OR  d.intervention_model=0
\end{lstlisting}
\vskip-1em
\caption{Example explanation view using $\projselect$, combining projection with the disjunction of 2 selection predicates. 
\label{fig:example_explanation_view_1}}
\end{subfigure}
\begin{subfigure}{0.9\linewidth}
\begin{lstlisting}
SELECT * FROM studies s
JOIN  facilities_studies fs  ON  s.nct_id=fs.nct_id
JOIN  facilities f  ON  fs.facility_id=f.facility_id
\end{lstlisting}
\vskip-1em
\caption{Example explanation view using $\join$, combining 2 important joins. 
\label{fig:example_explanation_view_2}}
\end{subfigure}
\end{figure}

\begin{table*}
\centering
\small
\caption{Explanation Metrics by Explanation Language for D6T1. \emph{Data retention (\%)} refers to the percentage of data, measured by the number of cells, that participate in the explanation views and are not perturbed during evaluation. \emptymethod\ corresponds to the empty explanation, i.e., where no information is retained.}
\label{tab:case_study_composite_langs}
\begin{tabular}{lccc}
\toprule
$\exlang$ & $\devi_\Delta(E)$ & \textbf{SQL Size} & \textbf{Data Retention (\%)} \\
\midrule
$\proj$  & 0.08   & 15   & 11.07  \\ 
$\join$  & 0.03   & 10   & 58.43 \\
$\select$  & 0.05 & 5 & 86.36 \\ 
$\projselect$  & 0.09   & 20   & 8.08   \\ 
$\joinproj$  & 0.08   & 25   & 11.07   \\ 
$\joinprojselect$  & 0.09   & 30  & 8.08   \\ 
\midrule
\emptymethod  & 0.25  & 0  & 0.0  \\ 
\bottomrule
\end{tabular}
\end{table*}

Here, we augment the Case Study of \Cref{sec:exp_case_study} with explanations using composite languages. 
\Cref{fig:example_explanation_view_1,fig:example_explanation_view_2} show two example explanation views returned by our framework. 
In \Cref{fig:example_explanation_view_1}, the explanation view projects onto the most important attributes of the relation \sql{designs}. Additionally, it applies the disjunction of two selection predicates which together select the most predictive study designs. The view indicates that knowledge about the study's design is highly predictive of its outcome, particularly for the selected design types. 
In \Cref{fig:example_explanation_view_2}, the view captures an important join path of length 2. The two joins correspond to two separate views in our language $\join$, but they are shown in a single view here. It indicates that the (potentially multiple) facilities where a study is conducted are important for its outcome.

\Cref{tab:case_study_composite_langs} illustrates how simple explanation views can yield low $\devi_\Delta$, below 0.1 for explanation languages such as $\join$ and $\proj$. The results also highlight the flexibility offered by the $\exlang$ choice. By combining projections, joins, and selections, one can balance size, data reduction, and expressivity.

\section{$\join$ Evaluation per Dataset-Task}

\begin{figure*}[b]
    \centering
    \caption{Evaluation for $\join$. For visualization purposes, tasks are separated into ``easy'' (left) and ``hard'' (right) ones, with separate $\devi_\Delta$ scales. The average (AVG) across all tasks is also shown. Below the dataset and task name, we indicate the explanation size $k^*$ and the total number of foreign-key constraints.}
    \label{fig:join-all}
    
    \begin{subfigure}[b]{0.25\textwidth}
        \centering
        \includegraphics[width=\textwidth]{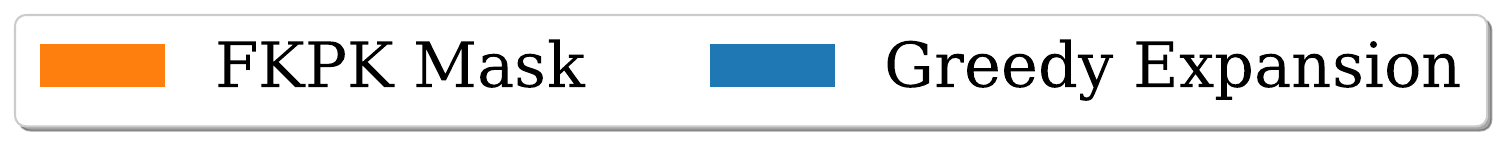}
    \end{subfigure}
    
    \begin{subfigure}[b]{\textwidth}
        \centering
        \includegraphics[width=0.8\textwidth]{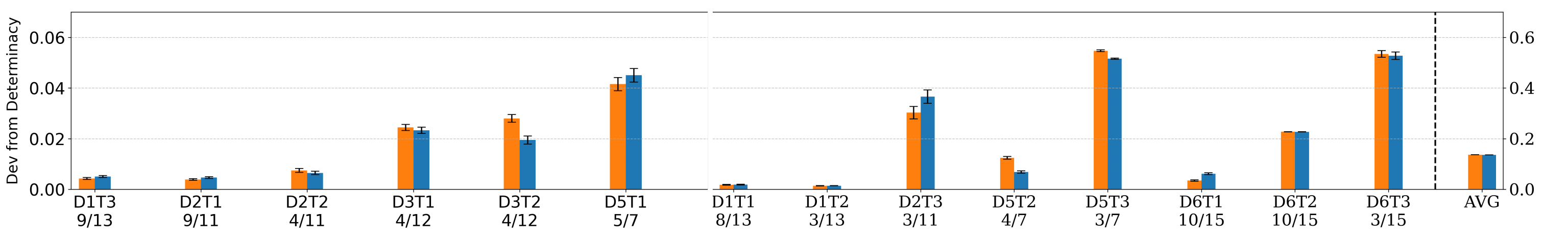}    
        \caption{Deviation from determinacy $\devi_\Delta$ for each database-task and database perturbation $\Delta^{\mathsf{uniform}}_{E}$.}
        \label{fig:join_dev_uniform}
    \end{subfigure}
        
    \begin{subfigure}[b]{\textwidth}
        \centering
        \includegraphics[width=0.8\textwidth]{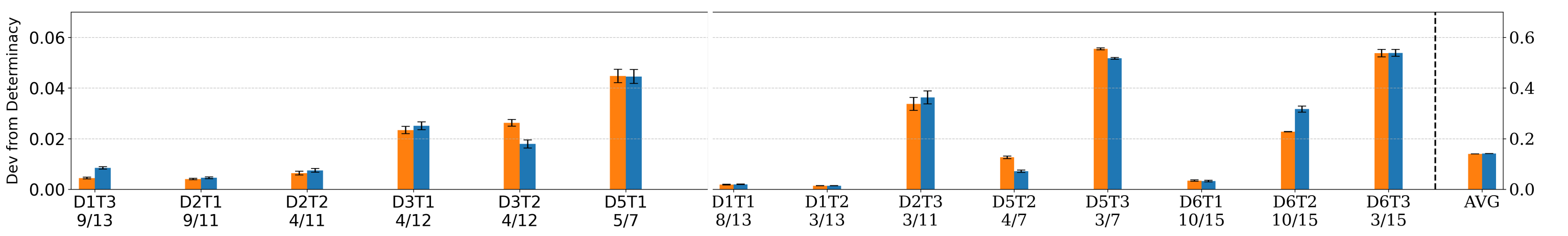}
        \caption{Deviation from determinacy $\devi_\Delta$ for each database-task and database perturbation $\Delta^{\mathsf{freq}}_{E}$.}
        \label{fig:join_dev_freq}
    \end{subfigure}
        
    \begin{subfigure}[b]{\textwidth}
        \centering
        \includegraphics[width=0.8\textwidth]{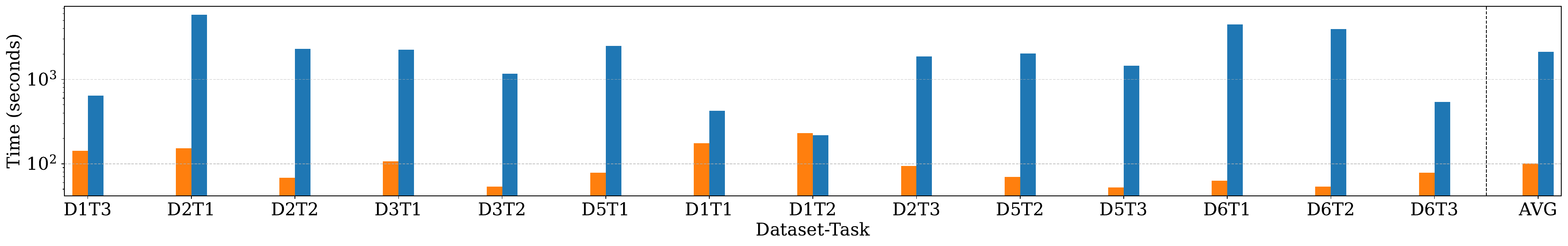}
        \caption{Time to obtain explanation for each database-task and $\Delta^{\mathsf{uniform}}_{E}$. Results for $\Delta^{\mathsf{freq}}_{E}$ are similar.}
        \label{fig:join_time}    
    \end{subfigure}
        
\end{figure*}

\Cref{fig:join-all} shows the results for $\join$ across all datasets and tasks, including averages (AVG).\Cref{fig:join_dev_uniform} and \Cref{fig:join_dev_freq} show $\devi_\Delta$ for two different perturbation strategies.

\end{document}